\documentclass[aps,preprint,pdftex]{revtex4}
\usepackage{amssymb}
\usepackage{amsmath}
\usepackage{graphicx}
\usepackage{bm}
\usepackage{color}
\usepackage{accents}
\usepackage{float}

\newcommand{\vect}[1]{\accentset{\rightharpoonup}{#1}}

\makeatletter

\newcommand{\Rmnum}[1]{\expandafter\@slowromancap\romannumeral #1@}
\makeatother

\begin{document}
\title{Neoclassical toroidal viscosity torque in tokamak edge pedestal induced by external resonant magnetic perturbation
}
\author{Xingting Yan}
\affiliation{CAS Key Laboratory of Geospace Environment and Department of Modern Physics \\ University of Science and Technology of China \\ Hefei, Anhui 230026, China}
\author{Ping Zhu}
\email[E-mail:~]{pzhu@ustc.edu.cn}
\affiliation{CAS Key Laboratory of Geospace Environment and Department of Modern Physics \\ University of Science and Technology of China \\ Hefei, Anhui 230026, China \\
\\
KTX Laboratory and Department of Modern Physics \\ University of Science and Technology of China \\ Hefei, Anhui 230026, China\\
 \\ Department of Engineering Physics \\  University of Wisconsin-Madison\\ Madison, Wisconsin 53706, USA}
\author{Youwen Sun}
\affiliation{CAS Institute of Plasma Physics \\ Hefei, Anhui 230031, China}

\date{\today}

\begin{abstract}
The characteristic profile and magnitude are predicted in theory for the neoclassical toroidal viscosity (NTV) torque induced by plasma response to the resonant magnetic perturbation (RMP) in a tokamak with an edge pedestal. For a low-$\beta$ equilibrium, the NTV torque is dominated by the toroidal component with the same dominant toroidal mode number of RMP. The NTV torque profile is found to be localized, whose peak location is determined by profiles of both the equilibrium temperature (pressure) and the plasma response. In general, the peak of NTV torque profile is found to track the pedestal position. The magnitude of NTV torque strongly depends on the $\beta$ value of pedestal, which suggests a more significant role of NTV torque in higher plasma $\beta$ regimes. For a fixed plasma $\beta$, decreasing density hence increasing temperature can also enhance the amplitude of NTV torque due to the reduced collisionality in the $1/\nu$ regime. Based on those findings, we identify the tokamak operation regimes where the significance of NTV torque in edge pedestal induced by RMP can approach those from other momentum sources such as the neutral beam injections.
\end{abstract}

\maketitle

\section{Introduction}
It is well known that plasma rotation plays critical roles in tokamak plasma dynamics, especially in the edge region~\cite{terry00a}. It is believed that plasma rotation with sufficient magnitude and shear is responsible for suppressing the edge turbulence, improving the edge transport, and triggering the L-H transition process~\cite{itoh88a,shaing89a,shaing90a}. Plasma rotation can also influence edge localized modes (ELMs). For example, in JT-60U experiments, type \Rmnum{1} ELM can change to high frequency grassy ELM when the counter plasma rotation is increased in the high plasma triangularity regime, while in low triangularity regime, complete ELM suppression can be obtained during the application of neutral beam injections (NBIs) \cite{Oyama}. \citet{Burrell_1} have found quiescent H-mode plasmas with strong edge rotation, where ELMs are replaced by edge harmonic oscillations (EHOs). 

On the other hand, it is found that ELMs can be mitigated or suppressed by the application of external resonant magnetic perturbations (RMPs) in various tokamaks~\cite{Evans, Burrell_2, Liang}. Yet how RMPs exactly suppress ELMs remains an open question~\cite{Evans_2015_PPCF, Nazikian_2015_PRL, Callen_CPTC}. As we know, RMP can affect plasma rotation by exerting resonant torques on the corresponding resonant surfaces, and non-resonant torque on plasma elsewhere due to effects from neoclassical toroidal viscosity (NTV) (i.e. NTV torque)~\cite{Shaing03, Park09}. Thus RMP may also influence edge pedestal instabilities like ELMs through its effects on edge plasma rotation. To explore such a possibility, we take the first step to evaluate the NTV torque in tokamak edge pedestal induced by RMPs. Generally speaking, any kind of non-axisymmetric magnetic perturbation (NAMP) can induce NTV torque. Once the NAMP field exists, it can produce a non-ambipolar diffusion of trapped particles, which gives rise to a radial current and the resulting NTV torque due to Lorentz force. 

Traditionally, the effects of NAMP are usually neglected in tokamaks because the typical amplitude of NAMP is only about $1\%$ of the inverse aspect ratio $\epsilon=a/R$ in tokamaks, where $a$ and $R$ are minor and major radius of a tokamak respectively. However, significant effects of NAMP on toroidal rotation have been observed in recent experiments, including the non-resonant magnetic braking or acceleration ~\cite{Zhu, Garofalo, Sabbagh, Hua, Sun_jet, Sun_jet_textor}. In addition, whereas the method of neutral beam injection (NBI) has been widely used to control plasma rotation in tokamak experiments, it may be insufficient to drive a high toroidal rotation in large-sized machine like ITER \cite{Rice}. The NAMPs induced by external coils may provide a desirable additional source of momentum input as well as a designed scheme for plasma rotation control. 

Over the last decade, theory predictions for different asymptotic collisionless regimes of NTV~\cite{Shaing03, Shaing08, Shaing09_1, Shaing09_2, Shaing09_3, Shaing09_4, Shaing10} and numerical models for NTV calculation~\cite{Kim_2012_POP, Logan_2013_POP, Liu_2013_POP, Wang_2014_POP, Satake_2013_NF, Honda_2014_NF} have been applied to explaining the observed effects of NAMPs in experiments, mostly in the core regions of tokamak plasmas~\cite{Zhu, Sabbagh, Garofalo, Hua, Sun_jet, Sun_jet_textor}, where the spatial variation of collisionality is weak in general. The situation however could be quite different in the edge pedestal region of an H-mode, where the collisionality can vary rapidly across the pedestal due to the steep gradient of pedestal temperature. One such example of equilibrium with an edge pedestal is shown in Fig.~\ref{fig_eq}. Whereas the collisionality is nearly a constant in the core region, the collisionality in the pedestal region spans at least $2$ orders of magnitude. Since NTV torque scales differently in different collisionality regimes, it is not clear how NTV torque may behave differently in the edge pedestal region.

The main objective of this work is to evaluate the NTV torque induced by external RMP and its significance on rotation in the edge pedestal. For the purpose of this evaluation, the initial coupling between the NIMROD code~\cite{Sovinec} and the NTVTOK code~\cite{Sun10_prl} has been developed to calculate the NTV torque induced by plasma response to RMP in a model tokamak with an edge pedestal. We evaluate the magnitude and profile of NTV torque and their scaling or dependence on pedestal properties and RMP amplitude. The NTV torque is found to be localized at edge, which is due to the combined effects of plasma response field and equilibrium temperature (pressure) profile. The magnitude of NTV torque is sensitive to the plasma $\beta$ and collisionality at the top of pedestal, in addition to the amplitudes of RMP and plasma response itself. Based on these findings, we are able to find the plasma regime and RMP amplitude for which the NTV torque in edge pedestal region can reach the same order of magnitude as that from the neutral beam injection.

The rest of the paper is organized as follows. In Sec.~\ref{sec:scheme} we introduce the coupling scheme developed between NIMROD and NTVTOK codes in this work. In Sec.~\ref{sec:prntv}, for a given calculated plasma response to RMP, the NTV torque from the plasma response is evaluated and its dependence on edge pedestal properties and RMP amplitude is described. Finally, we will give a summary and discussion in Sec.~\ref{sec:sum}.

\section{The coupling scheme for calculating NTV torque}
\label{sec:scheme}
In this work, we have developed an initial coupling scheme between the NIMROD and the NTVTOK codes to calculate the NTV torque due to non-axisymmetric magnetic perturbations in tokamaks. The particular models we used in the NIMROD and the NTVTOK codes, as well as the initial coupling scheme for these two codes will be briefly described in this section.

\subsection{Single-fluid MHD model in NIMROD code}

The single-fluid model implemented in NIMROD code is based on the following MHD equations \cite{Sovinec}:
\begin{equation}
\frac{\partial N}{\partial t}+\nabla\cdot (N\vect{u})=0,
\end{equation}
\begin{equation}
  MN(\frac{\partial\vect{u}}{\partial t}+\vect{u}\cdot\nabla\vect{u})=-\nabla p+\vect{J}\times\vect{B}-\nabla\cdot\overleftrightarrow{\Pi},
\end{equation}
\begin{equation}
  \frac{N}{\gamma-1}(\frac{\partial T}{\partial t}+\vect{u}\cdot\nabla T)=-p\nabla\cdot\vect{u}-\overleftrightarrow{\Pi}:\nabla\vect{u}-\nabla\cdot\vect{q},
\end{equation}
\begin{equation}\label{eq_Faraday}
\frac{\partial\vect{B}}{\partial t}=\nabla\times(\vect{u}\times\vect{B}-\eta\vect{J}),
\end{equation}
\begin{equation}
\mu_0\vect{J}=\nabla\times\vect{B},
\end{equation}
where $N$ is the number density, $M=m_i+m_e$, $m_i$ and $m_e$ the mass of ion and electron, $\vect{u}$ the fluid velocity, $p$ the plasma pressure, $T$ the plasma temperature, $\vect{B}$ the magnetic field, $\vect{J}$ the plasma current, $\eta$ the resistivity, $\gamma$ the adiabatic index, $\overleftrightarrow{\Pi}$ the anisotropic part of the pressure tensor, $\vect{q}$ the anisotropic heat flux, with 
$
\vect{q}=-N(\kappa_\parallel\nabla_\parallel T+\kappa_\perp\nabla_\perp T),
$
$\kappa_\parallel$ and $\kappa_\perp$ the parallel and perpendicular thermal diffusivities respectively. For any given external RMP field specified at tokamak boundary, the above equations are numerically solved as an initial-boundary problem using the NIMROD code to find the nonlinear plasma response $\delta\vect{B}$ as function of the cylindrical coordinates $(R,Z,\zeta)$ in a tokamak~\cite{Izzo_rmp, pzhu_aps}.

\subsection{NTV model in NTVTOK code}
\label{ntv_formula}

The evaluation of NTV torque density depends on the calculation of the perturbed distribution function $f_1$ and the perturbed magnetic field strength $\delta B$ along the distorted magnetic surfaces. NTV torque density profile can be written as~\cite{Shaing03, Sun11_nf, Sun_2013_NF}:
\begin{align}\label{eq_Tntv}
T_{NTV}(\hat{V})&=-\sum_j\left \langle R^2\nabla\zeta\cdot(\nabla \cdot\overleftrightarrow{\Pi}_j)\right\rangle_\psi \notag \\
&=-\sum_j\rho_jR_0^2\frac{\sqrt{\epsilon}q^2\omega_{tj}^2}{2\sqrt{2}\pi^{3/2}}\sum_n [\lambda_{1,n}(q\omega_E-q\omega_{*,j})-\lambda_{2,n}q\omega_{*T,j}],
\end{align}
where $\hat{V}$ represents the radial flux coordinate, and
\begin{equation}\label{eq_lambda}
\lambda_{l,n}=\frac{1}{2}\int_0^\infty I_{\kappa n}(x)(x-5/2)^{l-1}x^{5/2}e^{-x}dx,
\end{equation}
\begin{equation}\label{eq_Ikn}
I_{\kappa n}=\frac{|n\left\langle b_n\right\rangle_b|^2_{\kappa^2=0}}{\nu_d/(2\epsilon)}\int_0^1 4KF(I_1|\partial_{\kappa^2}f_{1n}|)^2d\kappa^2.
\end{equation}
Here $f_{1n}$ is the $n$-th Fourier component of perturbed distribution function, which can be obtained by solving the $n$-th component of bounce-averaged linearized drift kinetic equation (BDKE):
\begin{equation}\label{eq_BDKE_n}
I_1\left\langle L(f_{1n}) \right\rangle_b-iI_2f_{1n}-iI_3=0.
\end{equation}
The rest of definitions and the detailed solutions of the above BDKE implemented in the NTVTOK code, both analytical and numerical, can be found in~\cite{Shaing03, Shaing10, Sun11_nf, Sun_2013_NF}. The calculation of $b_n$, which is the $n$-th toroidal Fourier component of the perturbed magnetic field strength $\delta B$, is described in next subsection (Sec.~\ref{coupling_scheme}).

\subsection{Initial coupling scheme between NIMROD and NTVTOK}
\label{coupling_scheme}

The 3D perturbation of magnetic field $\delta\vect{B}(R,Z,\zeta)$ from NIMROD simulation can be used to evaluate the perturbed field strength $\delta B$ on the distorted magnetic surface and its Fourier component $b_n$ as in~\cite{Shaing03, Sun11_nf, Sun_2013_NF}
\begin{equation}
B=B_{eq}+\delta B,
\end{equation}
\begin{equation}\label{eq_bmn}
\delta B=-B_0\sum_{m,n}b_{mn}(\sqrt{\psi_p})e^{i(m\theta-n\zeta)}
=-B_0\sum_nb_n(\theta)e^{in\alpha},
\end{equation}
\begin{equation}
b_n(\theta)=\sum_mb_{mn} (\sqrt{\psi_p})e^{i(m-nq)\theta}, \alpha=q\theta-\zeta,
\end{equation}
where $B_{eq}$ is the equilibrium magnetic field strength, $B$ is the total magnetic field strength, and $B_0$ is the magnetic field strength at magnetic axis. $\delta B$ can be decomposed in two parts:
\begin{equation}
\delta B=\delta_E B+\delta_\xi B=\delta\vect{B}\cdot(\vect{B}/B)+\vect{\xi}\cdot\nabla B,
\end{equation}
$\vect{\xi}$ represents the displacement of flux surface, which can be evaluated from $\delta\vect{B}(R,Z,\zeta)$. Details of the calculation of $\delta B$ can be found in~\cite{Boozer_2006_POP, Park_2009_POP, Sun_jet, Sun_jet_textor, Sun_2015_PPCF}.

The coordinates adopted in NTV theory are Hamada coordinates $(\hat{V},\theta,\zeta)$, while those in NIMROD are cylindrical coordinates $(R,Z,\zeta)$, thus the coordinate transformation is necessary in the above calculation. In this work, for the first step, the coordinate $(\sqrt{\psi_p},\theta,\zeta)$ is adopted, where $\psi_p$ is the normalized poloidal magnetic flux, $\theta$ and $\zeta$ are evaluated using the geometric poloidal and toroidal angles, respectively.

To evaluate the radial profile of NTV torque, the equilibrium quantities required in Eq.~(\ref{eq_Tntv}) need to be flux functions. Since $\epsilon=\rho/R_0$ is no longer a flux function in toroidal geometry, where $\rho=\sqrt{\psi_T/\pi B_0}$, $\psi_T$ is the toroidal magnetic flux, $R_0$ the major radius at the magnetic axis, we adopt a $\theta$-averaged $\epsilon$, i.e.,
\begin{equation}
\epsilon=\int_0^{2\pi} \frac{\rho}{R_0} d\theta/\int_0^{2\pi} d\theta,
\end{equation}
and the integration is taken along the flux surface. Other equilibrium quantities, such as $q$, $T$ and $\omega_\zeta$, are flux functions so that can be used in the calculation directly.

\section{NTV torque induced by plasma response to RMP}
\label{sec:prntv}

\subsection{Plasma response to RMP}

\subsubsection{Equilibrium profiles}

\begin{figure}[ht]
  \includegraphics[width=0.57\textwidth,height=0.3\textheight]{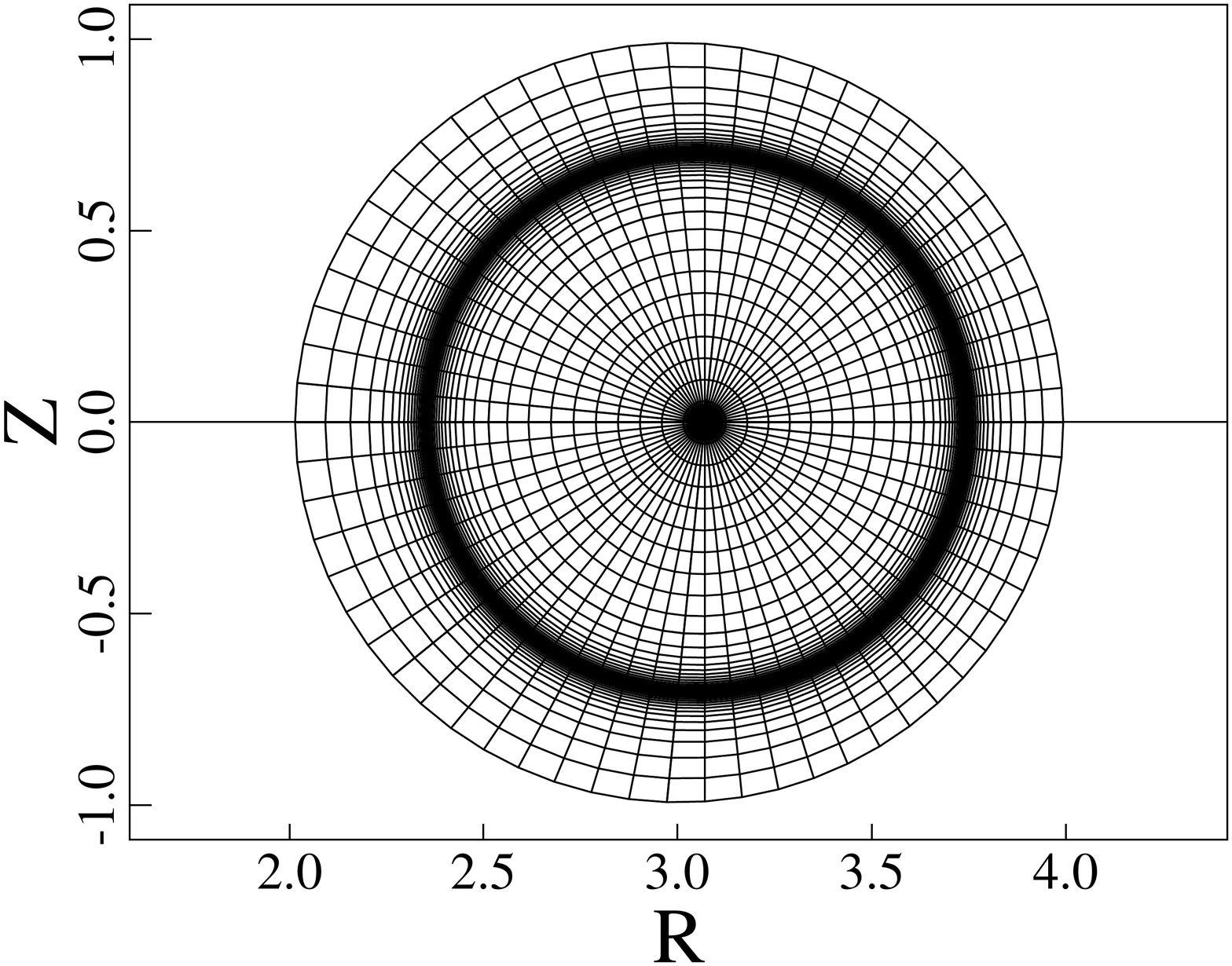}
  \put(-285,190){\textbf{(a)}}
  \vfill
  \includegraphics[width=0.6\textwidth,height=0.3\textheight]{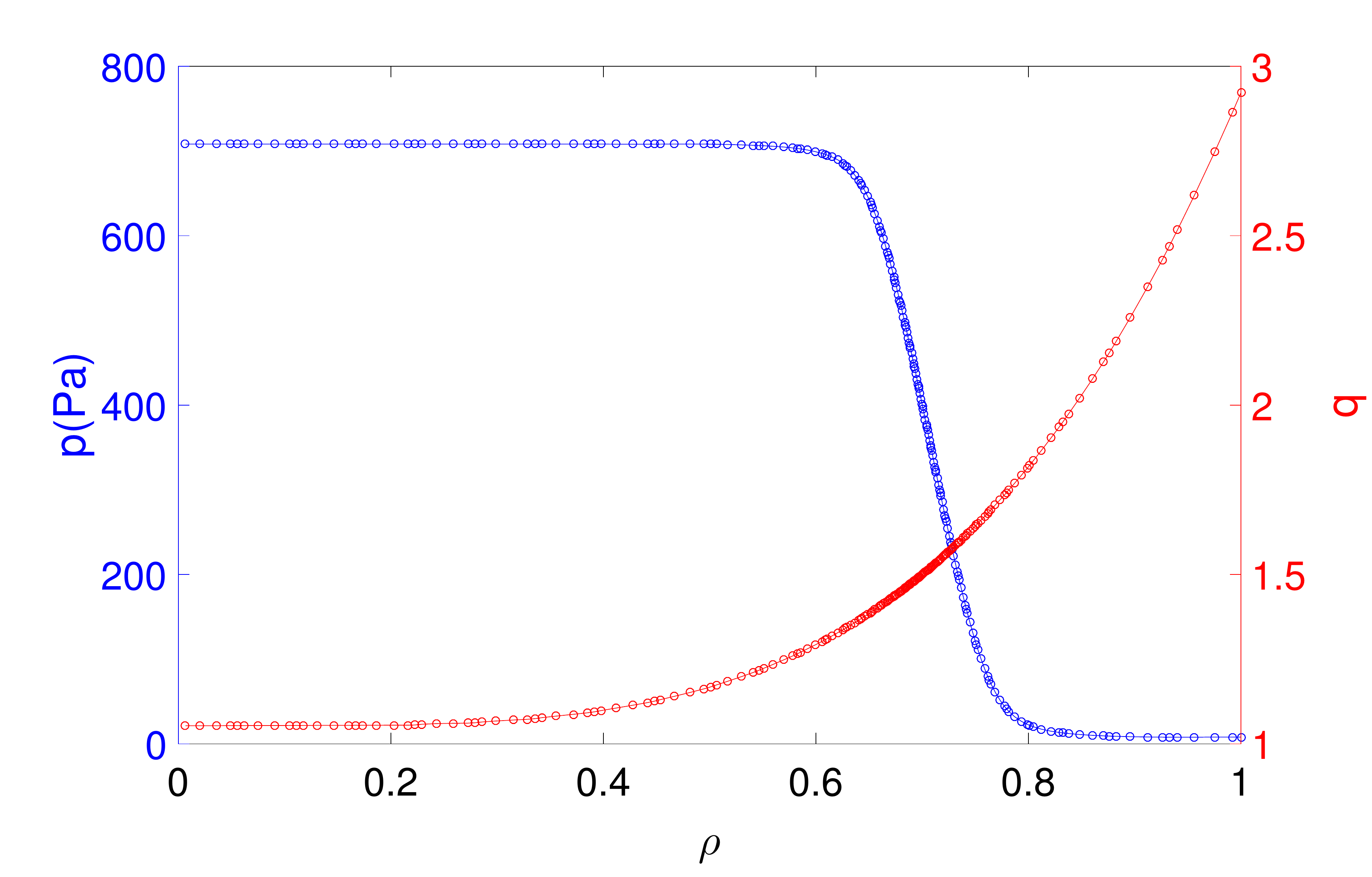}
  \put(-285,190){\textbf{(b)}}
  \vfill
  \includegraphics[width=0.6\textwidth,height=0.3\textheight]{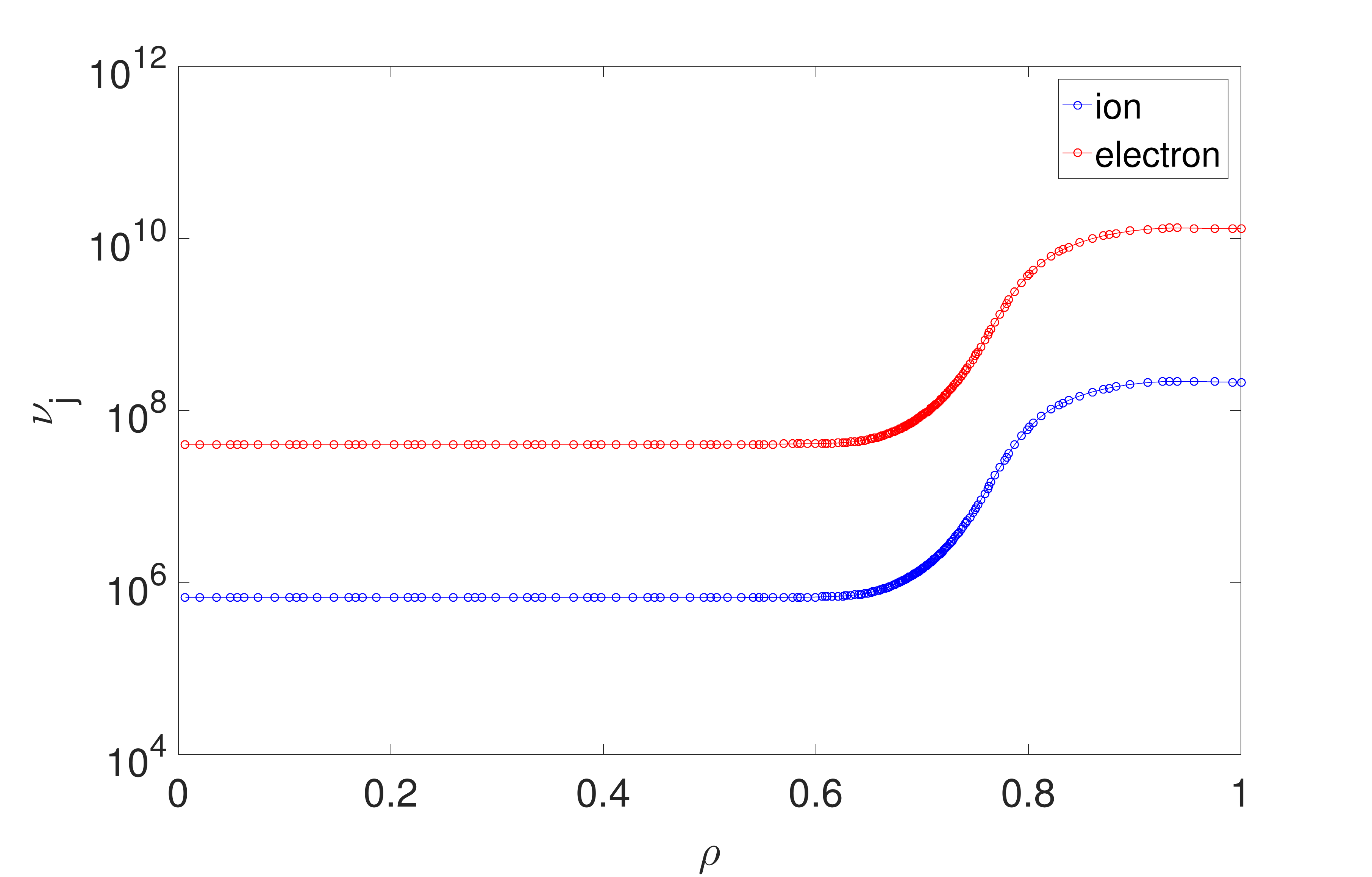}
  \put(-285,190){\textbf{(c)}}
  \caption{(a) Simulation domain and grid used in the study. Radial profiles as functions of minor radius $\rho$ for (b) pressure (blue line) and safety factor (red line), and (c) collisionalities of ion (blue line) and electron (red line).}
  \label{fig_eq}
\end{figure}

A static limiter-tokamak equilibrium with circular-shaped wall boundary is considered in this study (Fig.~\ref{fig_eq}a). The safety factor profile is specified as (Fig.~\ref{fig_eq}b)
\begin{equation}
q(\rho)=q_0\times [1+(\frac{q_a}{q_0}-1)\rho^4],
\end{equation}
where $q_0=1.05$, $q_a=3$, $q_0$ and $q_a$ are safety factors at magnetic axis and plasma boundary, respectively. Here and in the rest of the paper, $\rho$ represents the normalized minor radius, i.e. normalized $\sqrt{\psi_T/\pi B_0}$. The pressure profile is flat in the core region and has a steep gradient in the edge pedestal region (Fig.~\ref{fig_eq}b). Pressure profile follows the form:
\begin{equation}
p(\rho)=h_{ped}\times\tanh\frac{\rho_{ped}-\rho}{w_{ped}}+p_{ped},
\end{equation}
where $\rho_{ped}=0.7$, $w_{ped}=0.05$, $\mu_0h_{ped}=0.00044$, $\mu_0p_{ped}=0.00045$, $\rho_{ped}$, $w_{ped}$ and $h_{ped}$ indicate the position, width and height of the pedestal, respectively, and $p_{ped}$ is the pedestal pressure at $\rho_{ped}$. Other equilibrium parameters include: major radius $R_0=3m$, minor radius $a=1m$, number density $N=const.=10^{20}m^{-3}$, core ion temperature $T_{i,core}\approx 22eV$, $B_0=1T$, and the equilibrium toroidal flow is zero. To avoid complications from excitations of edge localized instabilities, a subcritical value of $\beta$ at top of pedestal ($\beta\approx\mu_0p(0)/B_0^2=0.088\%$) is chosen as the baseline case. The Lundquist number $S=\frac{a^2B_0}{R_0}\frac{1}{\sqrt{\mu_0NM}}\frac{\mu_0}{\eta}=2\times 10^4$, magnetic Prandtl number $Pr_m=1$. Assuming the same ion and electron temperatures, the collisionality profiles of ion and electron can be calculated. As mentioned in the introduction section, the collisionality changes by two orders of magnitude across the edge pedestal region (Fig.~\ref{fig_eq}c). The radial grid lines used in the NIMROD simulations in this work conform to the flux surfaces of equilibrium (Fig.~\ref{fig_eq}a). The radial grid is packed around the pedestal to obtain a high resolution of that region.

\subsubsection{Plasma response to RMP}\label{baseline}

NIMROD calculations find the equilibrium stable for all toroidal modes in the resistive time scale ($t\sim 1ms$) in absence of any external drive, such as RMP. To calculate plasma response, RMP is applied as a fixed boundary condition at the tokamak wall, similar to the previous NIMROD calculations~\cite{Izzo_rmp, pzhu_aps}. We choose the helicity of RMP as $(m,n)=(2,1)$, where $m$, $n$ are poloidal and toroidal mode numbers, and the amplitude of RMP as $10^{-4}T$. Six toroidal modes ($n=0-5$) are included in the nonlinear NIMROD simulations of the plasma response. 

\begin{figure}[ht]
  \includegraphics[width=0.5\textwidth,height=0.27\textheight]{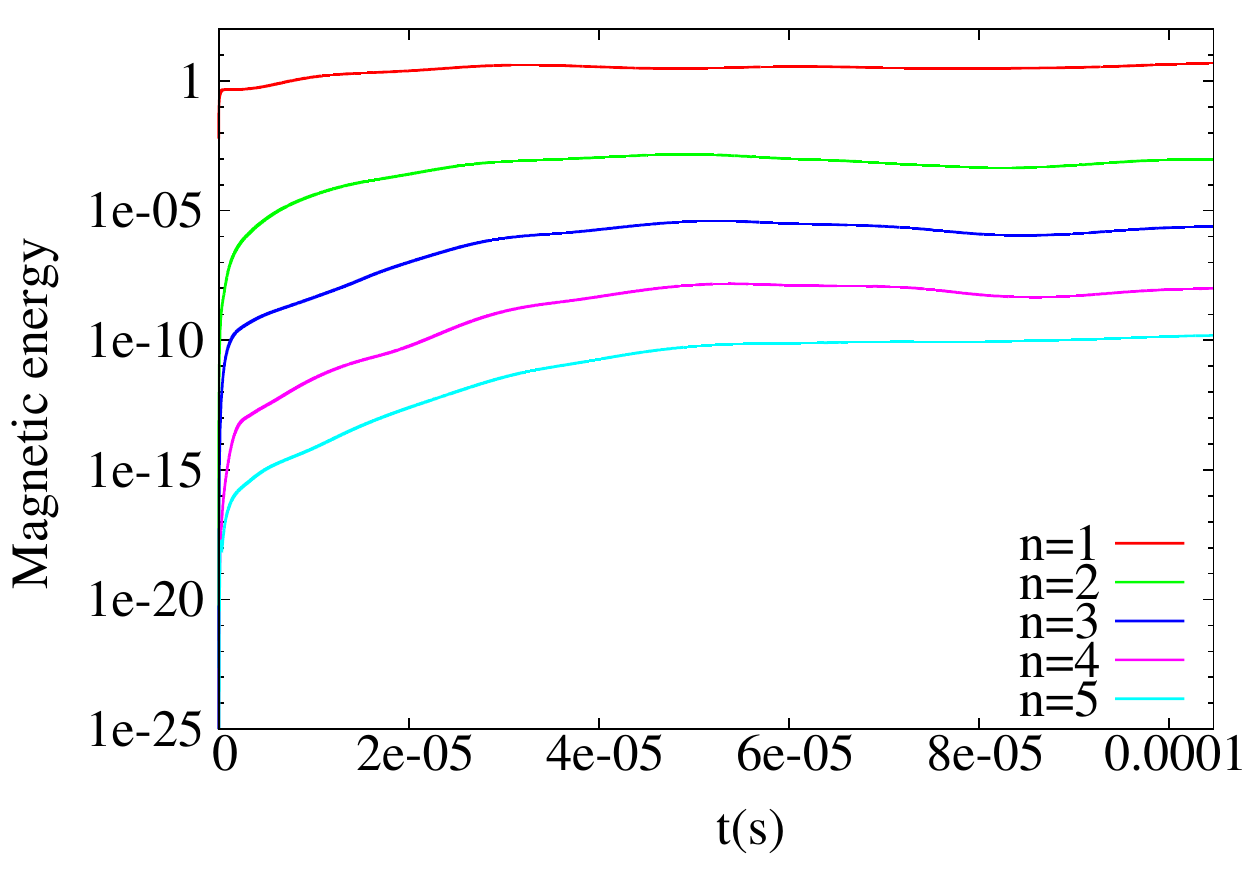}
  \put(-220,180){\textbf{(a)}}
  \hspace{0.65cm}\hfill
  \includegraphics[width=0.46\textwidth,height=0.27\textheight]{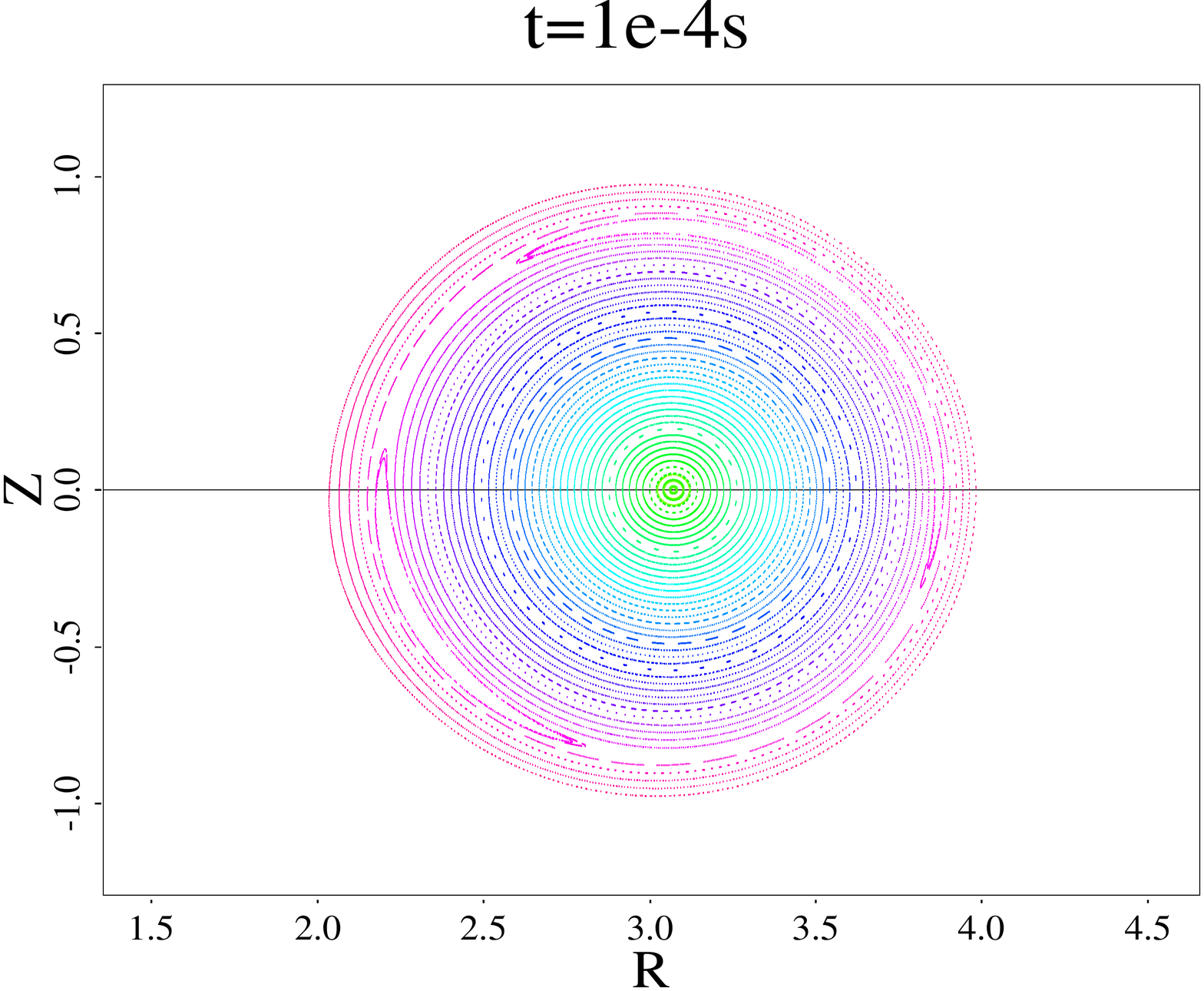}
  \put(-220,180){\textbf{(b)}}
  \vfill
  \includegraphics[width=0.55\textwidth,height=0.27\textheight]{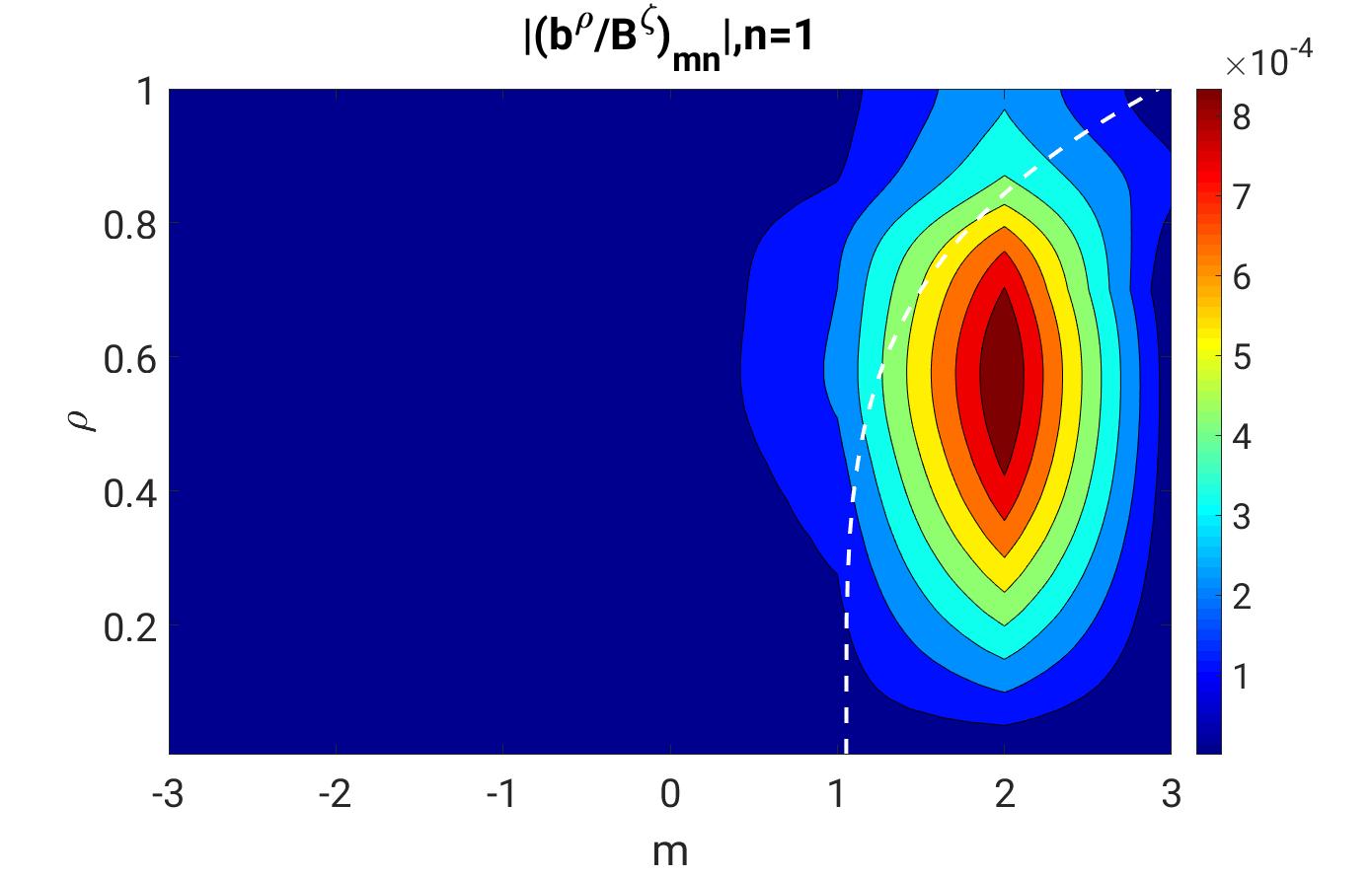}
  \put(-250,170){\textbf{(c)}}
  \includegraphics[width=0.55\textwidth,height=0.27\textheight]{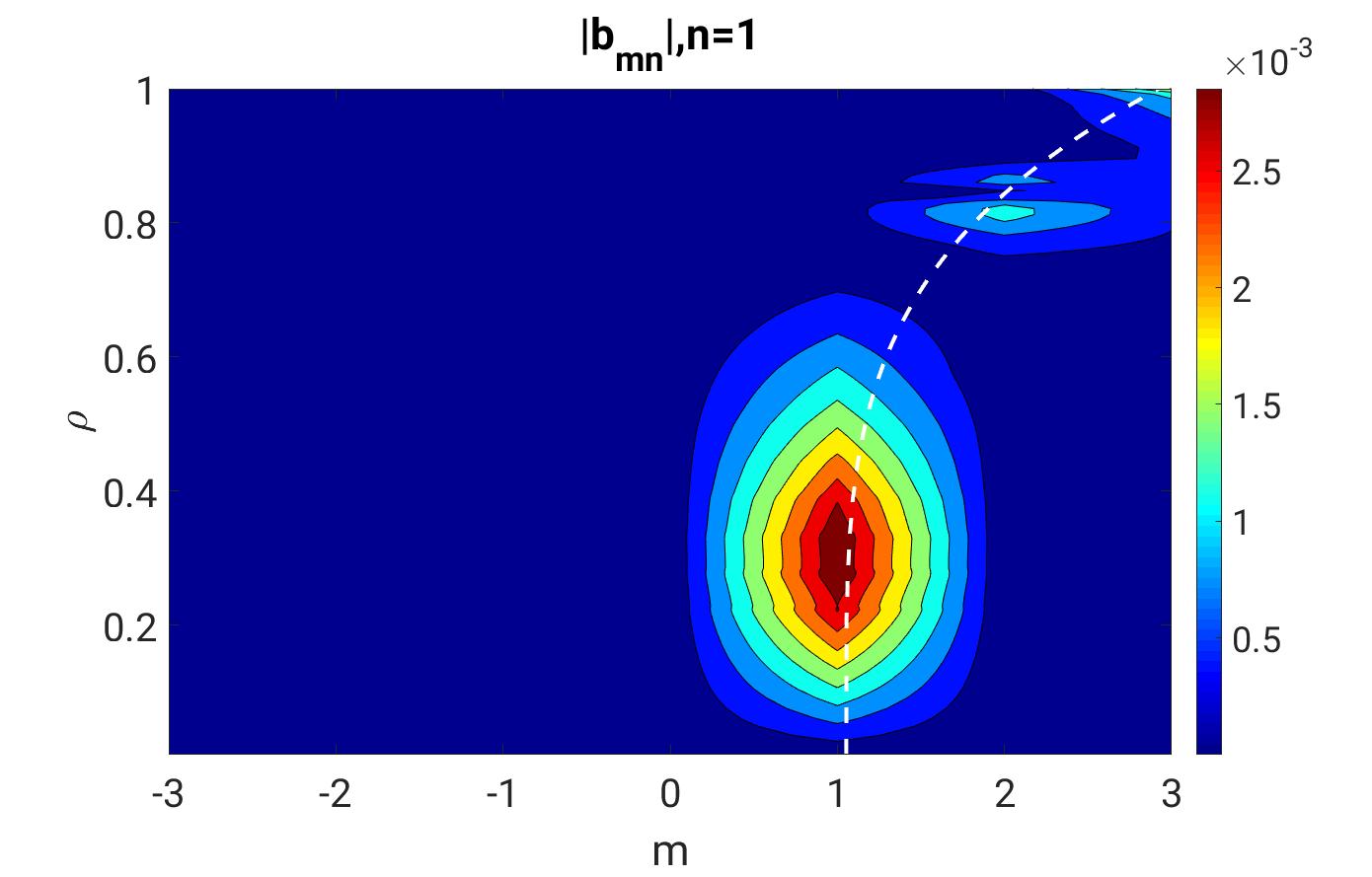}
  \put(-250,170){\textbf{(d)}}
  \vfill
  \includegraphics[width=0.55\textwidth,height=0.27\textheight]{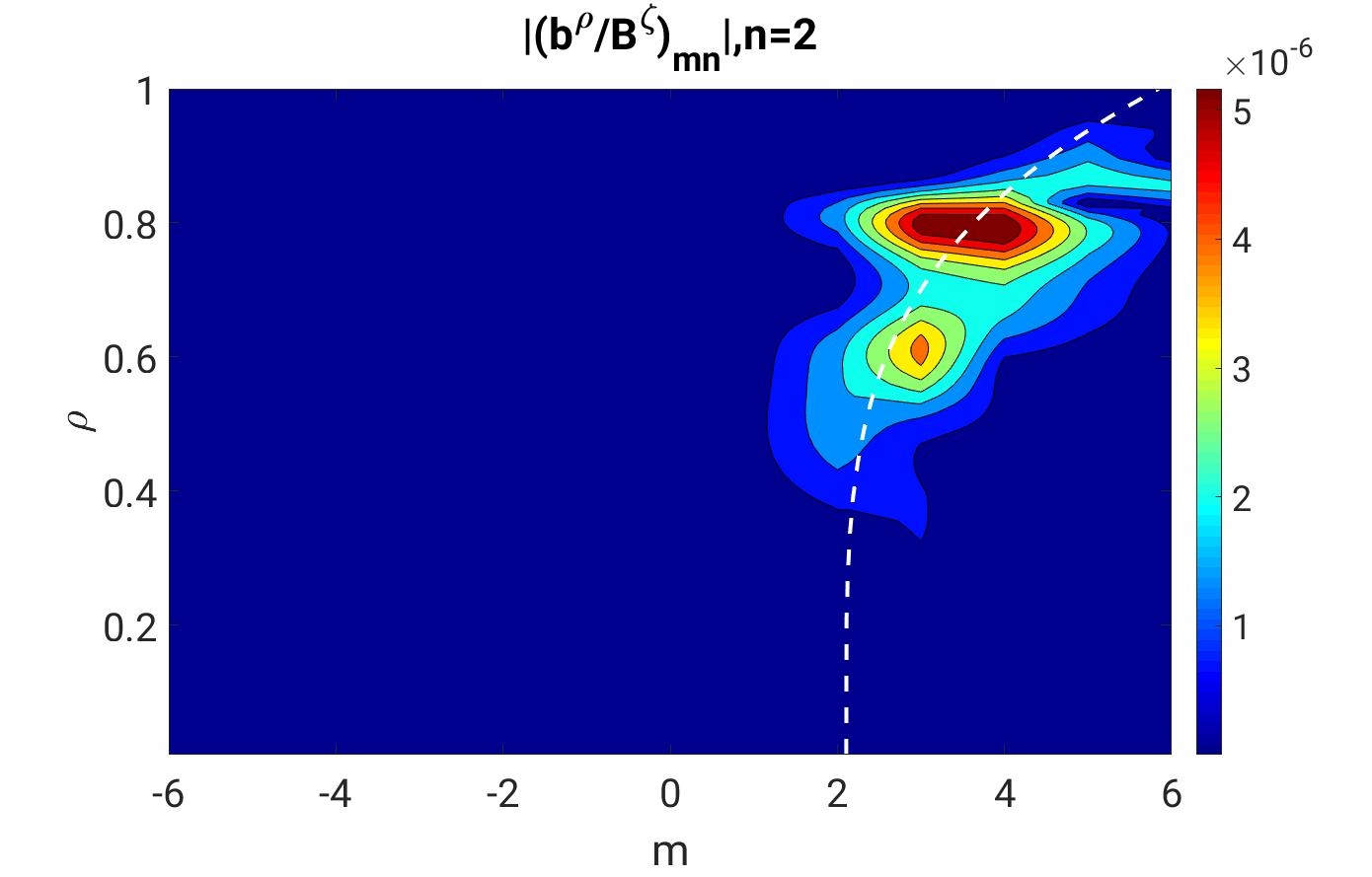}
  \put(-250,170){\textbf{(e)}}
  \includegraphics[width=0.55\textwidth,height=0.27\textheight]{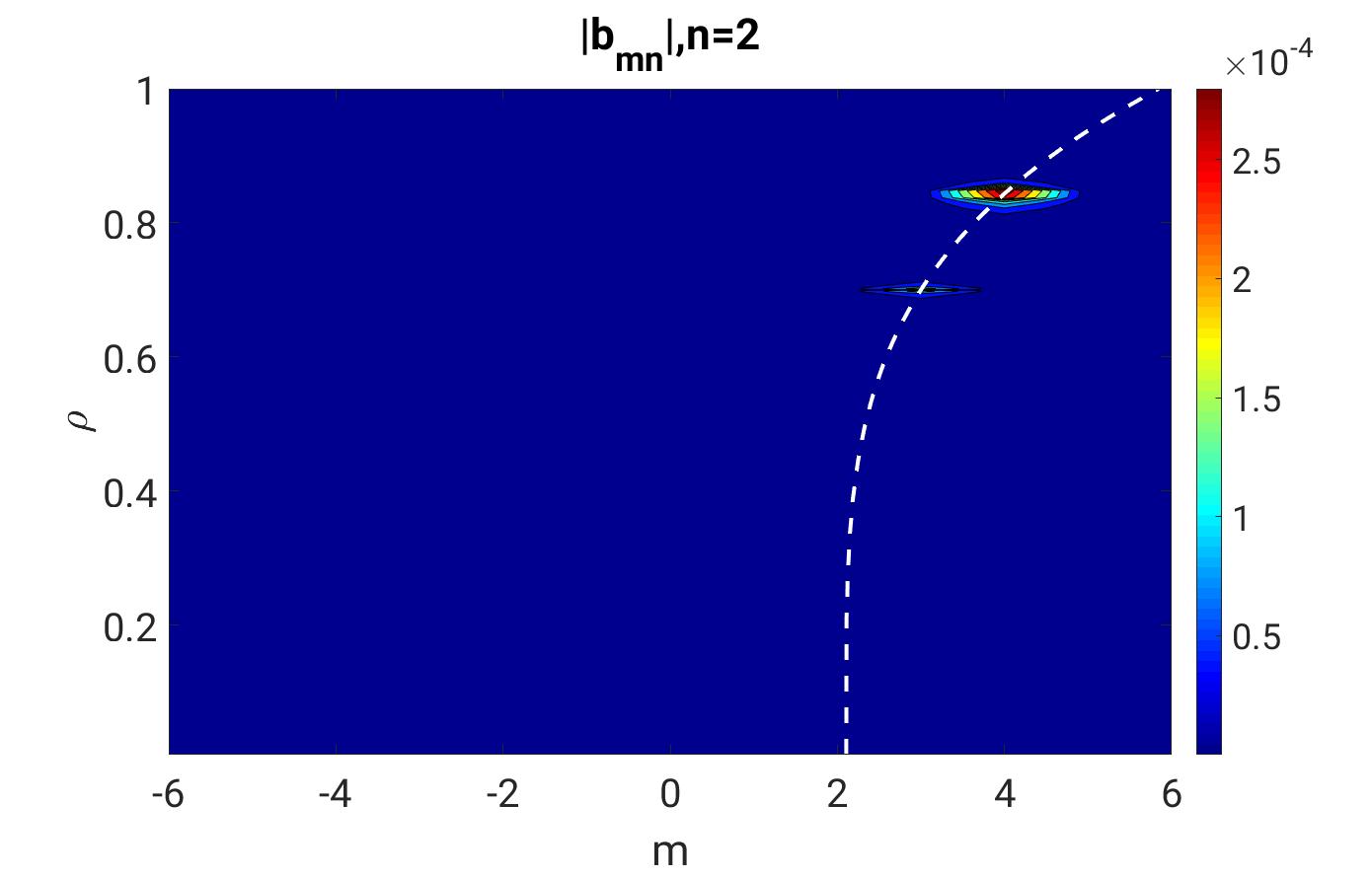}
  \put(-250,170){\textbf{(f)}}
  \caption{(a) Time evolution of perturbed magnetic energies of different toroidal components. (b) Poincare plot at the saturation phase ($t=10^{-4}s$). Contours of (c) radial component of plasma response $|(b^{\rho}/B^{\zeta})_{mn}|$ and (d) perturbed magnetic field strength $|b_{mn}|$ for the $n=1$ component; (e) and (f) are the corresponding contours for the $n=2$ component.}
  \label{fig_base_res}
\end{figure}

From the time evolution of magnetic energies of different toroidal modes, we find that the amplitude of $n=1$ mode is dominant among all the toroidal components, and the saturation level decreases with toroidal mode number $n$ (Fig.~\ref{fig_base_res}a). After the RMP is applied, the perturbed energy saturates rapidly. The Poincare plot at the saturation phase ($t=10^{-4}s$) indicates that $(2,1)$ island appears at the corresponding resonant surface (Fig.~\ref{fig_base_res}b). For the $n=1$ component of the plasma response, the contour plot of $|(b^\rho/B^\zeta)_{mn}|$ on $\rho-m$ plane shows that the amplitude of $m=2$ component is dominant, and there is also a significant non-resonant part in the response field, i.e. the perturbed field is not localized at the resonant surface (Fig.~\ref{fig_base_res}c). Here $b^\rho=\delta\vect{B}\cdot\nabla\psi_p/|\nabla\psi_p|$, $B^\zeta=\vect{B}\cdot\nabla\zeta$. Nonetheless, the dominant helicity of plasma response is the same as that of RMP, i.e. $(m,n)=(2,1)$.

In contrast, the spectrum of the perturbed field strength $|b_{mn}|$, which is defined in Eq.~(\ref{eq_bmn}), is mainly localized around resonant surfaces (Fig.~\ref{fig_base_res}d). The dominant poloidal mode is $m=1$, which is induced by the distortion of flux surfaces near the $(1,1)$ surface in the core region. Such a distortion of flux surface due to magnetic field perturbation makes the major contribution to the perturbed field strength $|b_{mn}|$ and the resulting NTV torque.

The amplitudes of $n=2$ component of $|(b^\rho/B^\zeta)_{mn}|$ and $|b_{mn}|$ (Figs.~\ref{fig_base_res}e-f) are much smaller than those of $n=1$ component (Figs.~\ref{fig_base_res}c-d), as also can be inferred from the smaller saturation level of $n=2$ component of perturbed energy in Fig.~\ref{fig_base_res}a. The $|b_{mn}|$ profile is much more localized and peaked around resonant surfaces than $n=1$ component, which can induce a more localized contribution to the NTV torque profile, as will be shown in next subsection (Sec.~\ref{prof_ntv}).

\subsection{Profiles of NTV torque}\label{prof_ntv}

\begin{figure}[ht]
  \includegraphics[width=0.49\textwidth,height=0.25\textheight]{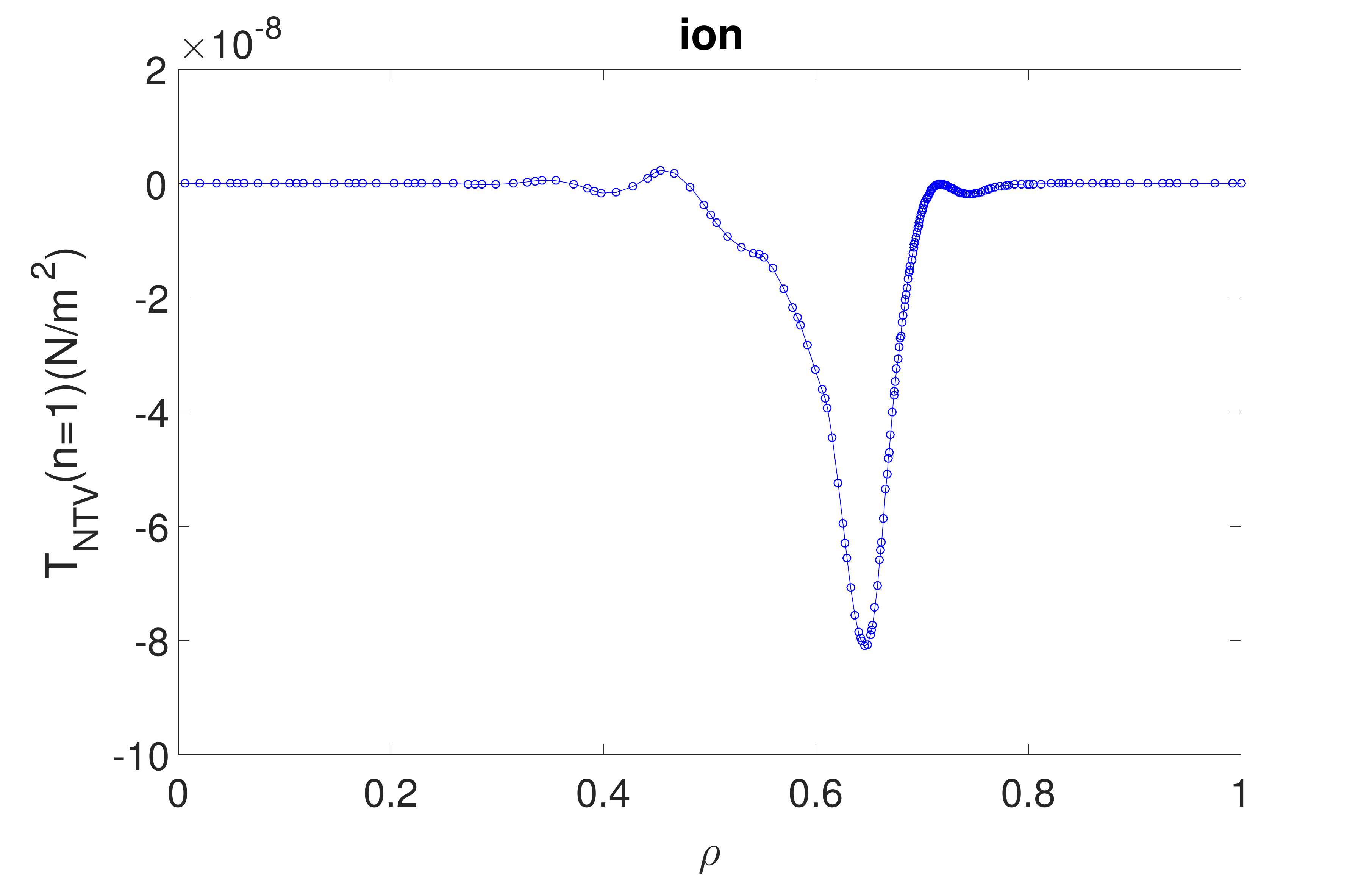}
  \put(-230,160){\textbf{(a)}}
  \includegraphics[width=0.49\textwidth,height=0.25\textheight]{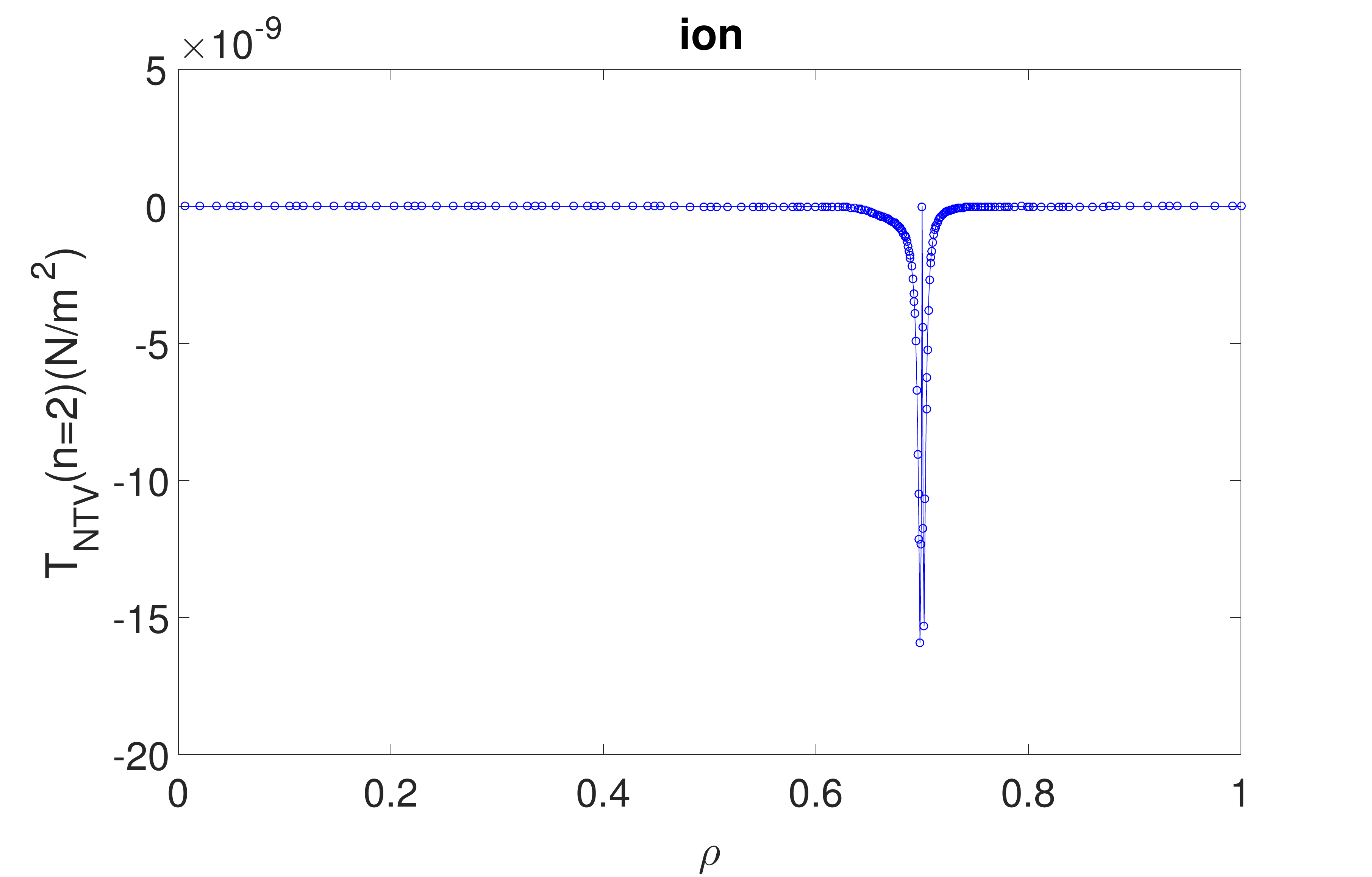}
  \put(-230,160){\textbf{(b)}}
  \vfill
  \includegraphics[width=0.49\textwidth,height=0.25\textheight]{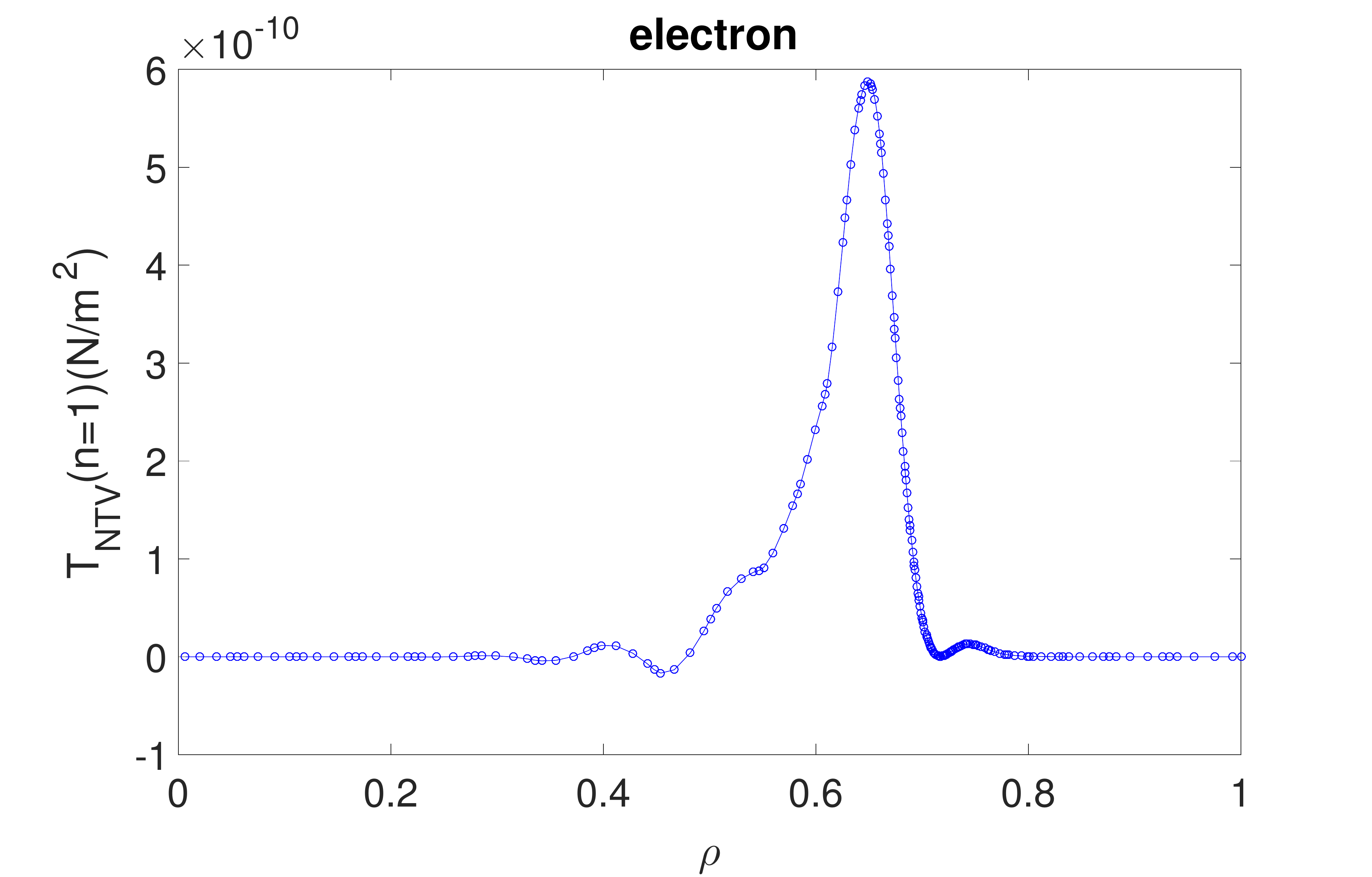}
  \put(-230,160){\textbf{(c)}}
  \includegraphics[width=0.49\textwidth,height=0.25\textheight]{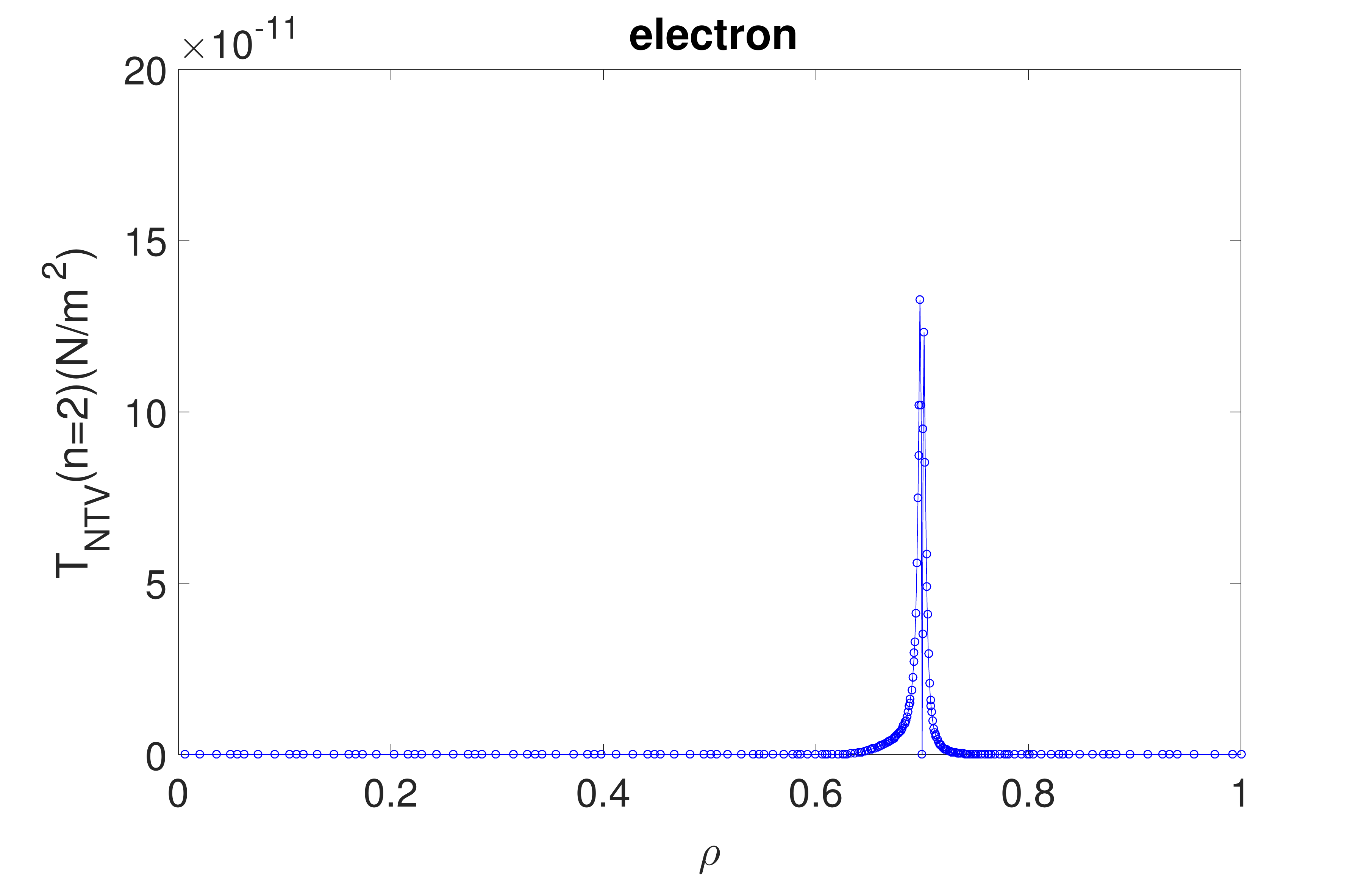}
  \put(-230,160){\textbf{(d)}}
  \vfill
  \caption{Radial profiles of $T_{NTV}$ as functions of minor radius $\rho$ for: (a) ion $T_{NTV}$ induced by the $n=1$ component of plasma response; (b) ion $T_{NTV}$ induced by the $n=2$ component of plasma response; (c) electron $T_{NTV}$ induced by the $n=1$ component of plasma response; (d) electron $T_{NTV}$ induced by the $n=2$ component of plasma response.}
  \label{fig_Tntv_baseline}
\end{figure}

The NTV torque profiles for both ions and electrons are localized radially, and the ion and electron NTV torques are in opposite directions for the case considered here (Fig.~\ref{fig_Tntv_baseline}). We find that the amplitude of ion NTV torque is much larger than that of electron NTV torque, however, electron NTV torque can usually become comparable to ion NTV torque or even dominant in low collisionality regimes. The magnitudes of NTV torque induced by the $n=1$ component of plasma response (Figs.~\ref{fig_Tntv_baseline}a and \ref{fig_Tntv_baseline}c) are much larger than those induced by the $n=2$ component of plasma response (Figs.~\ref{fig_Tntv_baseline}b and \ref{fig_Tntv_baseline}d), which is a consequence of the dominance of the $n=1$ component of plasma response as shown in the previous subsection (Sec.~\ref{baseline}). Because the $n=2$ component of plasma response is more localized, its contribution to NTV torque is also more localized than that from the $n=1$ component of plasma response. For the convenience of statement, the results shown in Fig.~\ref{fig_Tntv_baseline}, together with the corresponding equilibrium and plasma response shown in Figs.~\ref{fig_eq}-\ref{fig_base_res}, are referred to as the baseline case in this paper.

\begin{figure}[ht]
  \includegraphics[width=0.49\textwidth,height=0.25\textheight]{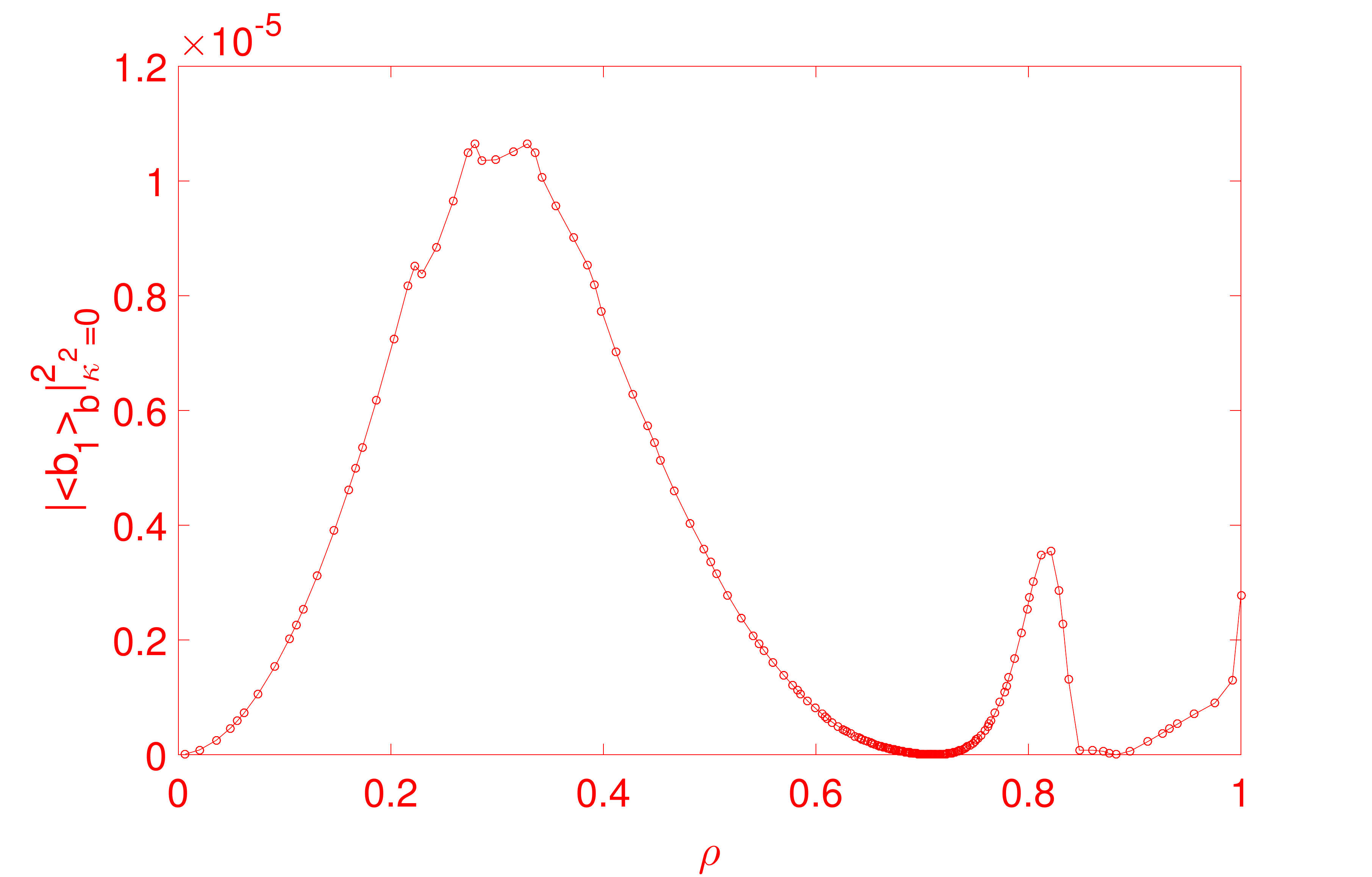}
  \put(-230,160){\textbf{(a)}}
  \includegraphics[width=0.49\textwidth,height=0.25\textheight]{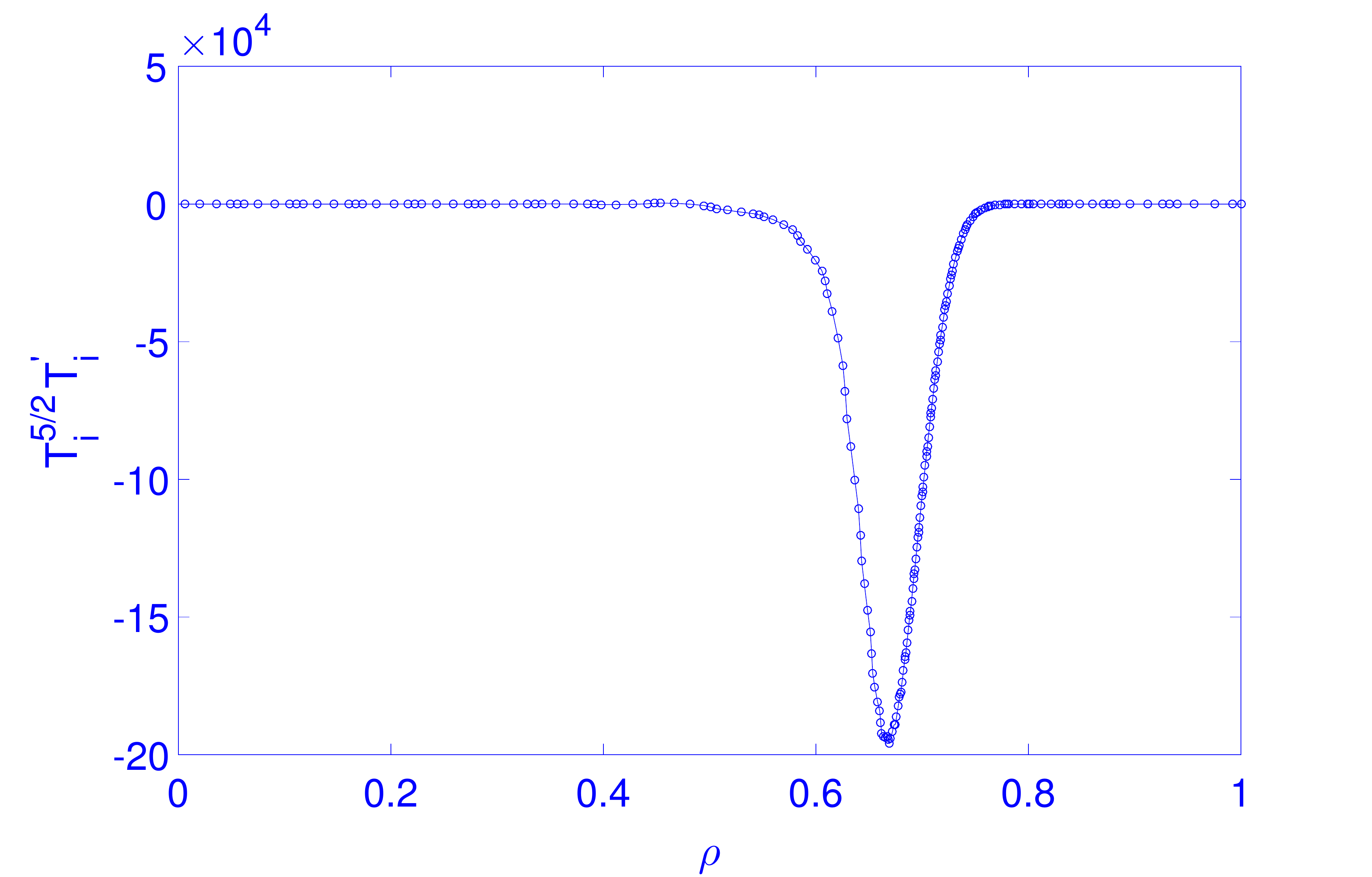}
  \put(-230,160){\textbf{(b)}}
  \vfill
  \includegraphics[width=0.49\textwidth,height=0.25\textheight]{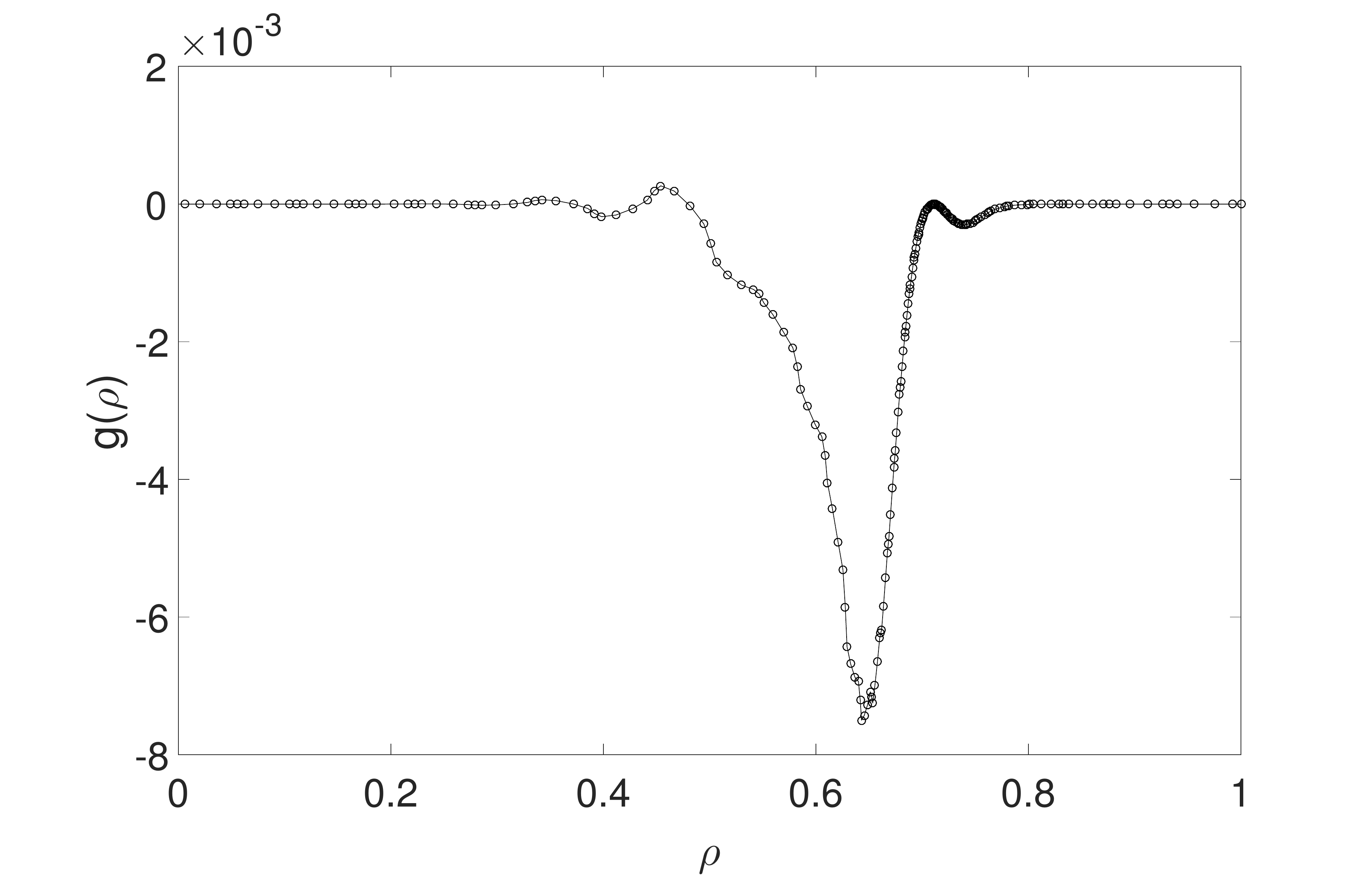}
  \put(-230,160){\textbf{(c)}}
  \includegraphics[width=0.49\textwidth,height=0.25\textheight]{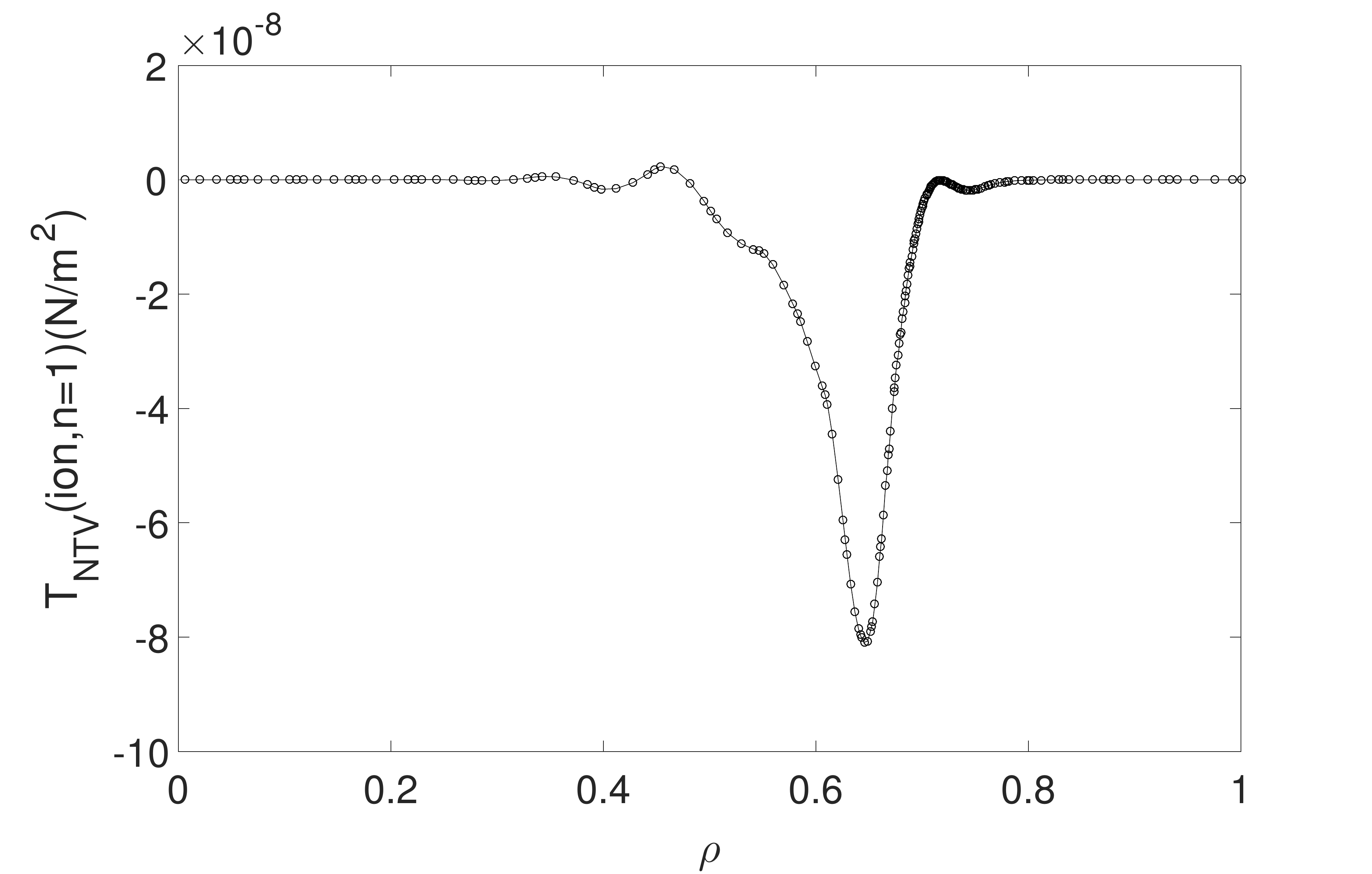}
  \put(-230,160){\textbf{(d)}}
  \vfill
  \caption{Radial profiles as functions of minor radius $\rho$ for: (a) bounce-averaged $n=1$ component of plasma response $|\left\langle b_1 \right\rangle_b|_{\kappa^2=0}^2$; (b) $T_i^{5/2}T_i^{'}$, where $T_i$ is the ion temperature, prime denotes the derivative with respect to $\rho$; (c) the function $g(\rho)$; (d) ion $T_{NTV}$ induced by the $n=1$ component of plasma response.}
  \label{fig_Tntv_approx}
\end{figure}

The NTV torque profile is determined by the combined effects of both plasma response and equilibrium profiles, and in particular, the profiles of $b_n$, $q\omega_E$, $q\omega_*$, and $q\omega_{*T}$ in Eqs.~(\ref{eq_Tntv})-(\ref{eq_Ikn}). These profiles are dependent on the perturbed magnetic field strength, plasma toroidal rotation, density, and temperature. In the case we study in this paper, toroidal rotation is zero ($\omega_\zeta=0$) and density is constant, therefore we can obtain the following reduced formula for NTV torque profile:
\begin{equation}\label{eq_Tntv_approx}
T_{NTV}(ion,n=1)\propto \frac{q}{\rho}\epsilon^{3/2}|\left\langle b_1\right\rangle_b|^2_{\kappa^2=0}T_i^{5/2}T_i^{'}\equiv g(\rho),
\end{equation}
where the prime denotes the derivative with respect to $\rho$. Only the ion NTV torque induced by the $n=1$ component of plasma response is considered above, which are both dominant in the baseline case. Comparing the profiles of $g(\rho)$ and $T_{NTV}$ in Figs.~\ref{fig_Tntv_approx}c-d, we find $g(\rho)$ a good approximation to the shape of $T_{NTV}$ profile. As illustrated in Eq.~(\ref{eq_Tntv_approx}), it is the combined effects of plasma response (i.e. $|\left\langle b_1 \right\rangle_b|_{\kappa^2=0}^2$ in Fig.~\ref{fig_Tntv_approx}a) and equilibrium profiles (i.e. $T_i^{5/2}T_i^{'}$ in Fig.~\ref{fig_Tntv_approx}b) that determine the NTV torque profile, even though Eq.~(\ref{eq_Tntv_approx}) is based on a simplified scenario considered in this study ($\omega_\zeta=0$, $N=const.$). Thus Eq.~(\ref{eq_Tntv_approx}) suggests an experimental scheme for achieving a large magnitude of $T_{NTV}$ through controlling equilibrium and plasma response profiles in experiments. In another word, sufficiently large perturbed field strength, temperature and temperature gradients are required to collocate in the same region in order to generate a maximal level of NTV torque.

\subsection{Pedestal effects on NTV torque}

Apparently, NTV torque depends on the equilibrium properties, not only directly, but also indirectly through plasma response. In order to investigate the direct dependence of NTV torque on the equilibrium profiles, we use a 'fixed plasma response' in this section, where the spectrum of perturbed magnetic field strength $b_{mn}$ as a function of $\sqrt{\psi_p}$ is kept the same as that in the baseline case, even as the equilibrium profiles are varied.

\begin{figure}[ht]
  \includegraphics[width=0.6\textwidth,height=0.3\textheight]{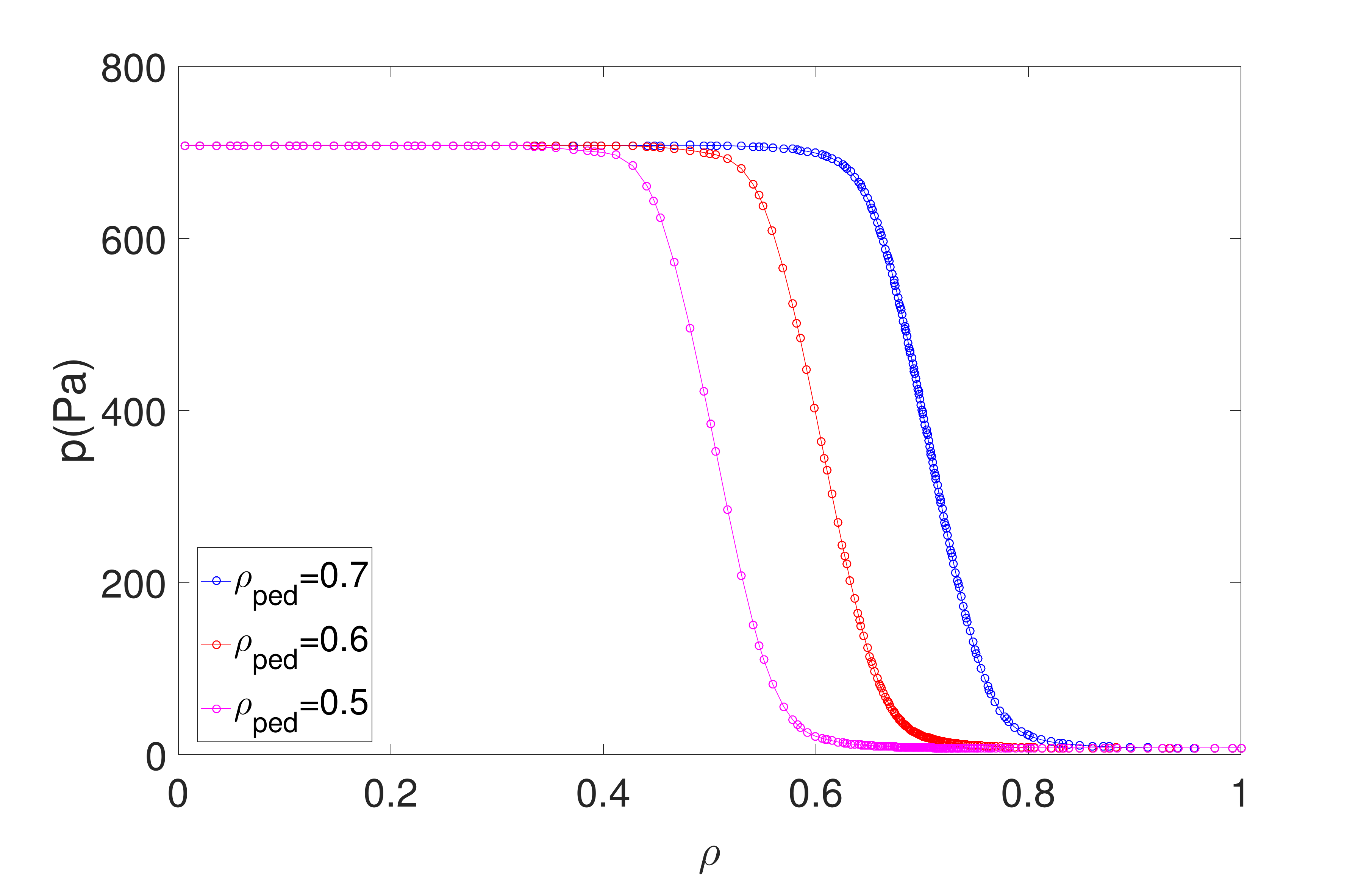}
  \put(-290,180){\textbf{(a)}}
  \vfill
  \includegraphics[width=0.6\textwidth,height=0.3\textheight]{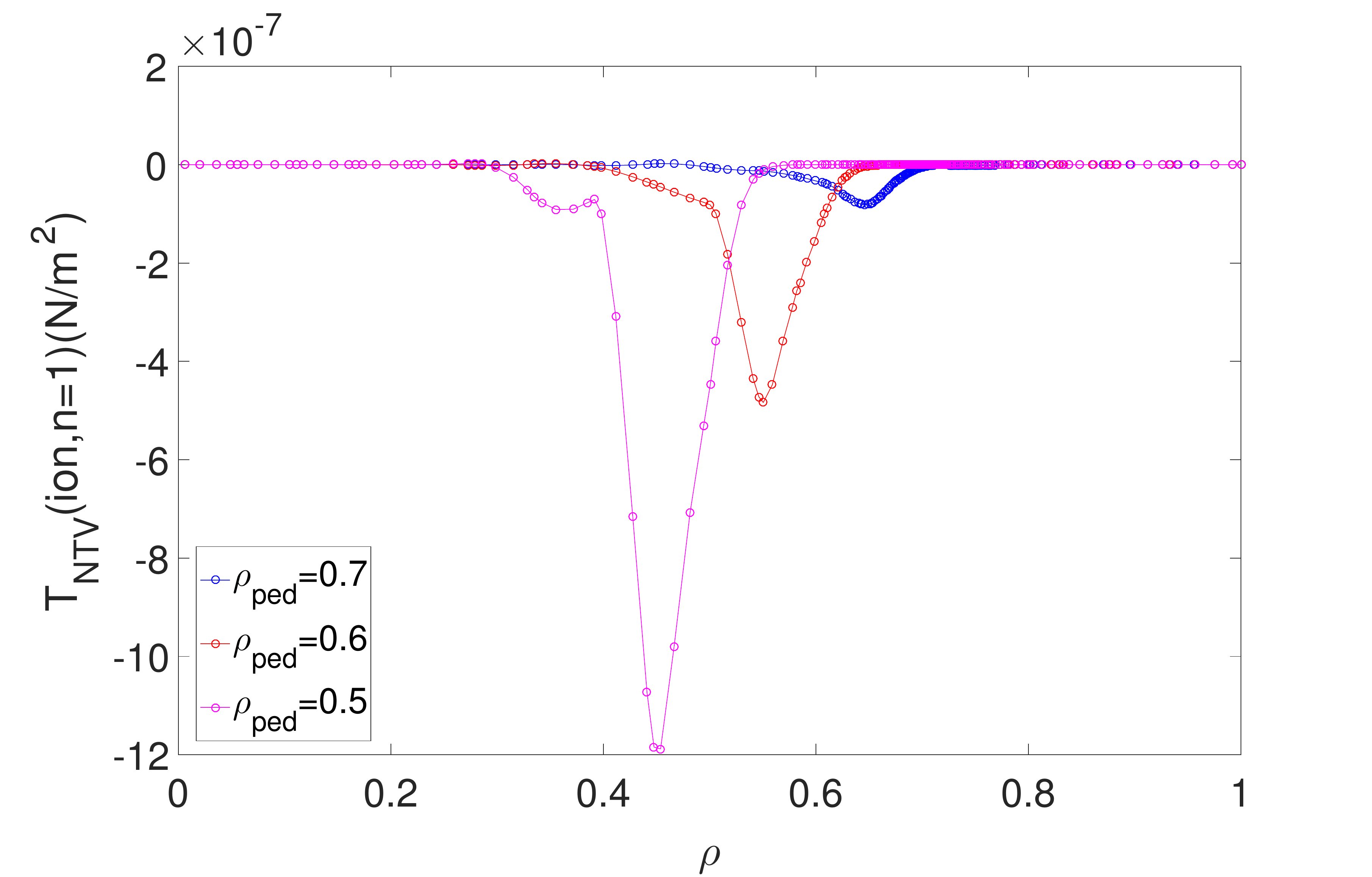}
  \put(-290,180){\textbf{(b)}}
  \vfill
  \includegraphics[width=0.6\textwidth,height=0.3\textheight]{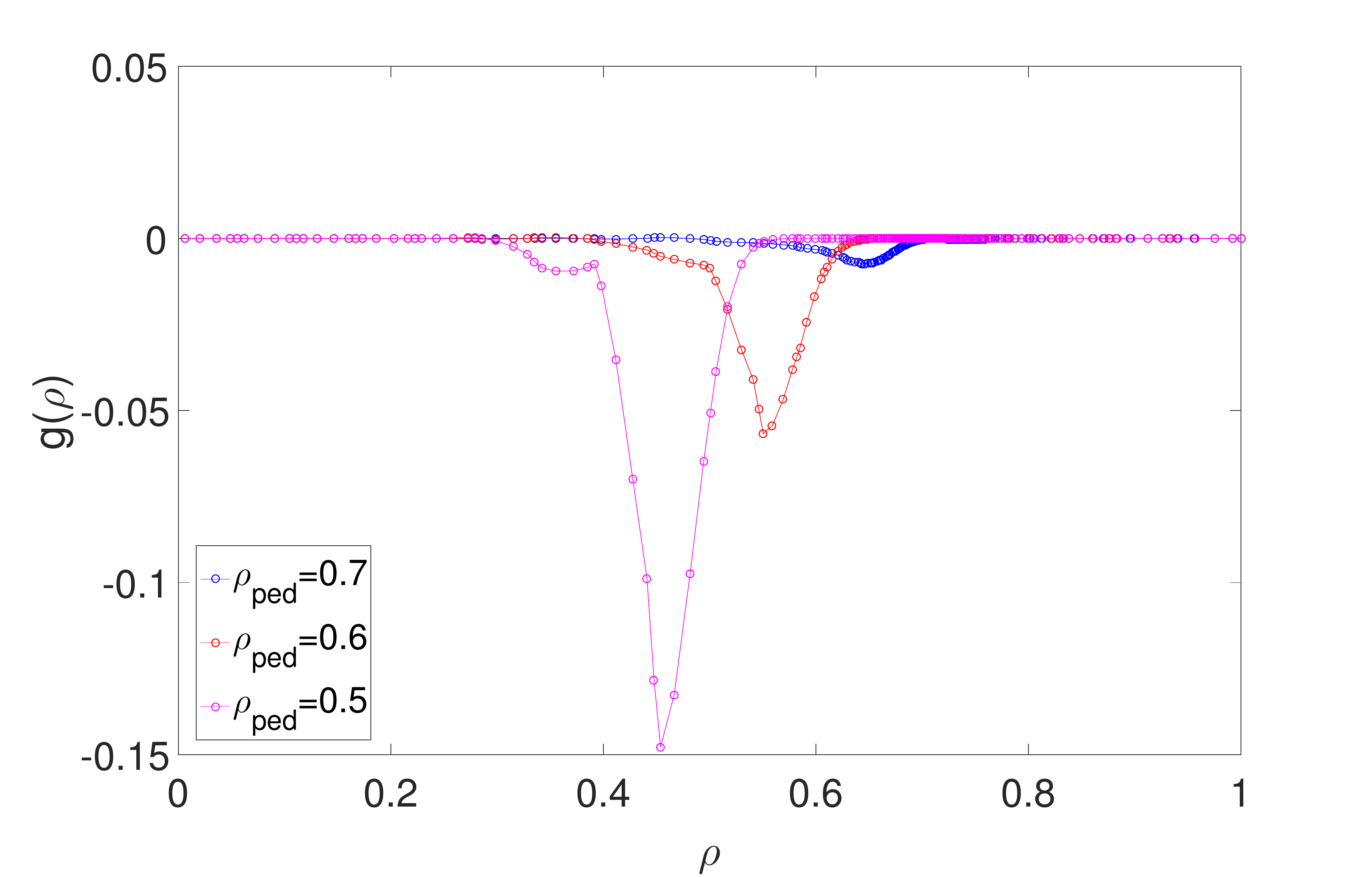}
  \put(-290,180){\textbf{(c)}}
  \caption{Radial profiles as functions of minor radius $\rho$ for: (a) equilibrium pressure; (b) ion $T_{NTV}$ induced by a fixed $n=1$ component of plasma response; (c) the function $g(\rho)$ of equilibria with different pedestal locations. $\rho_{ped}$ indicates the pedestal location.}
  \label{fig_Tntv_ped}
\end{figure}

The first key equilibrium property is the pedestal location. The combined effects of plasma response and equilibrium profiles can be stronger if the peaked locations of perturbed field strength and pedestal gradient are closer to each other, as suggested in Eq.~(\ref{eq_Tntv_approx}). Because the $(m,n)=(1,1)$ component of perturbed field strength is dominant, which represents the distortion of flux surfaces in the core region, NTV torque is mainly induced by this component. Indeed, as the pedestal moves inward (Fig.~\ref{fig_Tntv_ped}a) towards the locations of peaked value of perturbed field strength, the peak of NTV torque profile is found to track the pedestal position, and the amplitude of NTV torque increases significantly (Fig.~\ref{fig_Tntv_ped}b). The function $g(\rho)$ in Eq.~(\ref{eq_Tntv_approx}) can also give a good approximation to the shape of NTV torque profile in these cases (Fig.~\ref{fig_Tntv_ped}c), which captures the dependence of NTV torque on the relative locations of equilibrium and plasma response profiles.

\begin{figure}[ht]
  \includegraphics[width=0.6\textwidth,height=0.3\textheight]{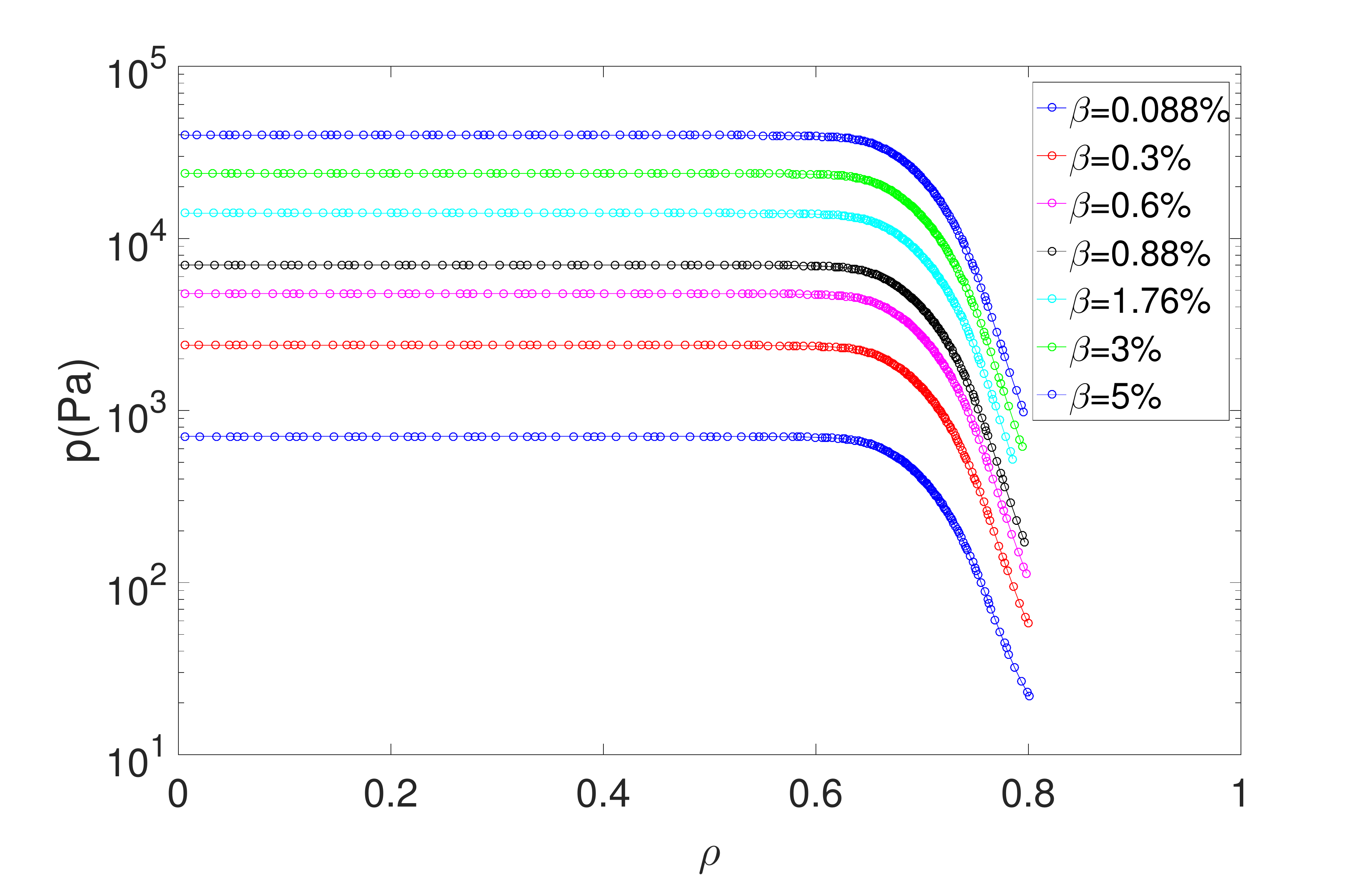}
  \put(-290,180){\textbf{(a)}}
  \vfill
  \includegraphics[width=0.6\textwidth,height=0.3\textheight]{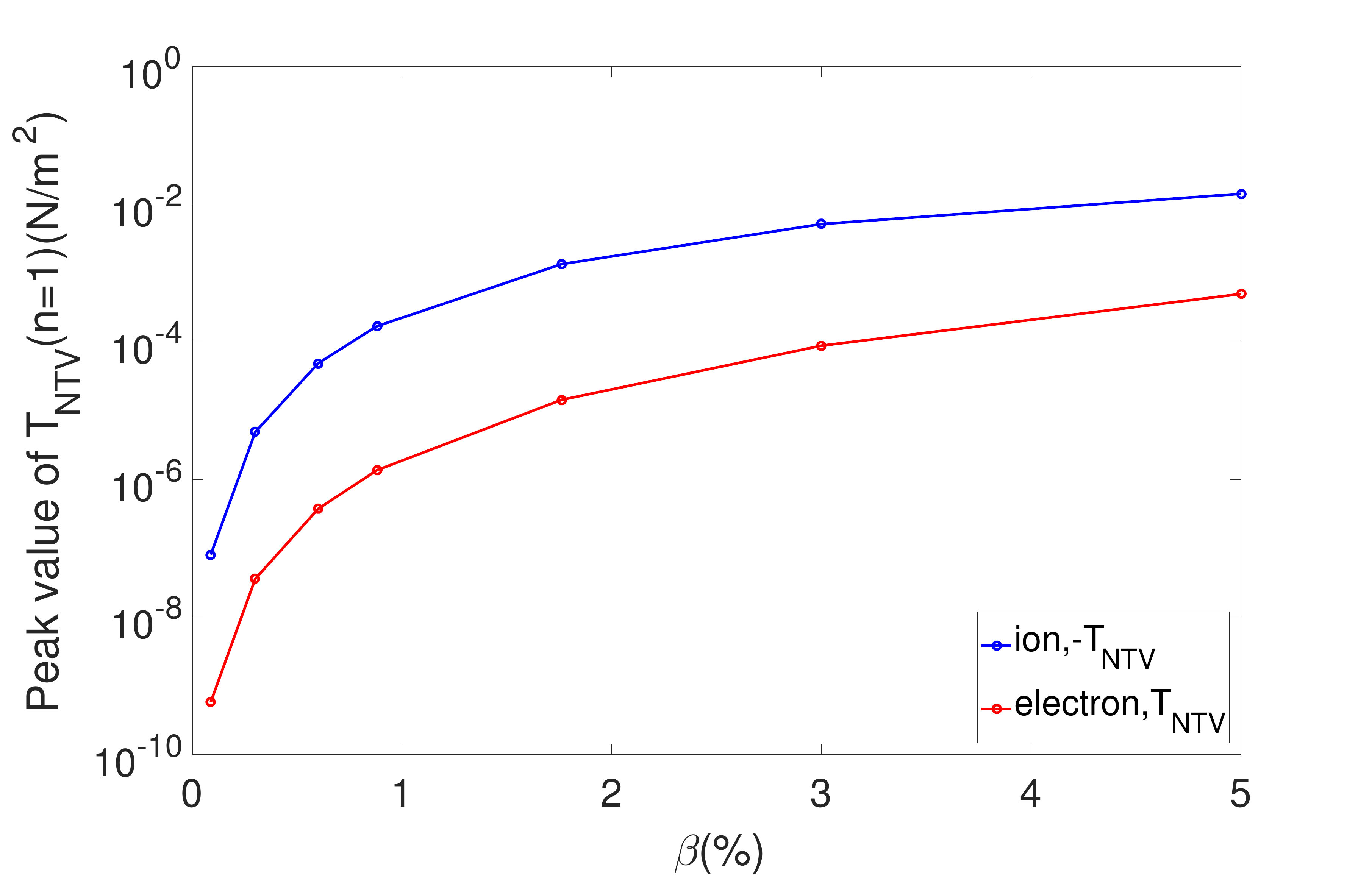}
  \put(-290,180){\textbf{(b)}}
  \caption{(a) Radial profiles of equilibrium pressure as functions of minor radius $\rho$ for different plasma $\beta$. (b) Peak values of ion $T_{NTV}$ (blue line) and electron $T_{NTV}$ (red line) as functions of $\beta$ induced by a fixed $n=1$ component of plasma response.}
  \label{fig_Tntv_beta}
\end{figure}

Another key edge pedestal parameter is the plasma $\beta$ at top of pedestal. With the variation of $\beta$, the pressure (temperature) gradient also changes (Fig.~\ref{fig_Tntv_beta}a). We find that the peak value of $T_{NTV}$ can be quite sensitive to $\beta$ (Fig.~\ref{fig_Tntv_beta}b), in low-$\beta$ regime, when $\beta$ increases by one order of magnitude, the amplitude of $T_{NTV}$ increases by three orders of magnitude. However, we should note that the dependence of NTV torque on $\beta$ becomes weaker in high-$\beta$ regime, which means that merely increasing $\beta$ may not be sufficient for increasing NTV torque in high-$\beta$ regime.

\begin{figure}[ht]
  \includegraphics[width=0.6\textwidth,height=0.29\textheight]{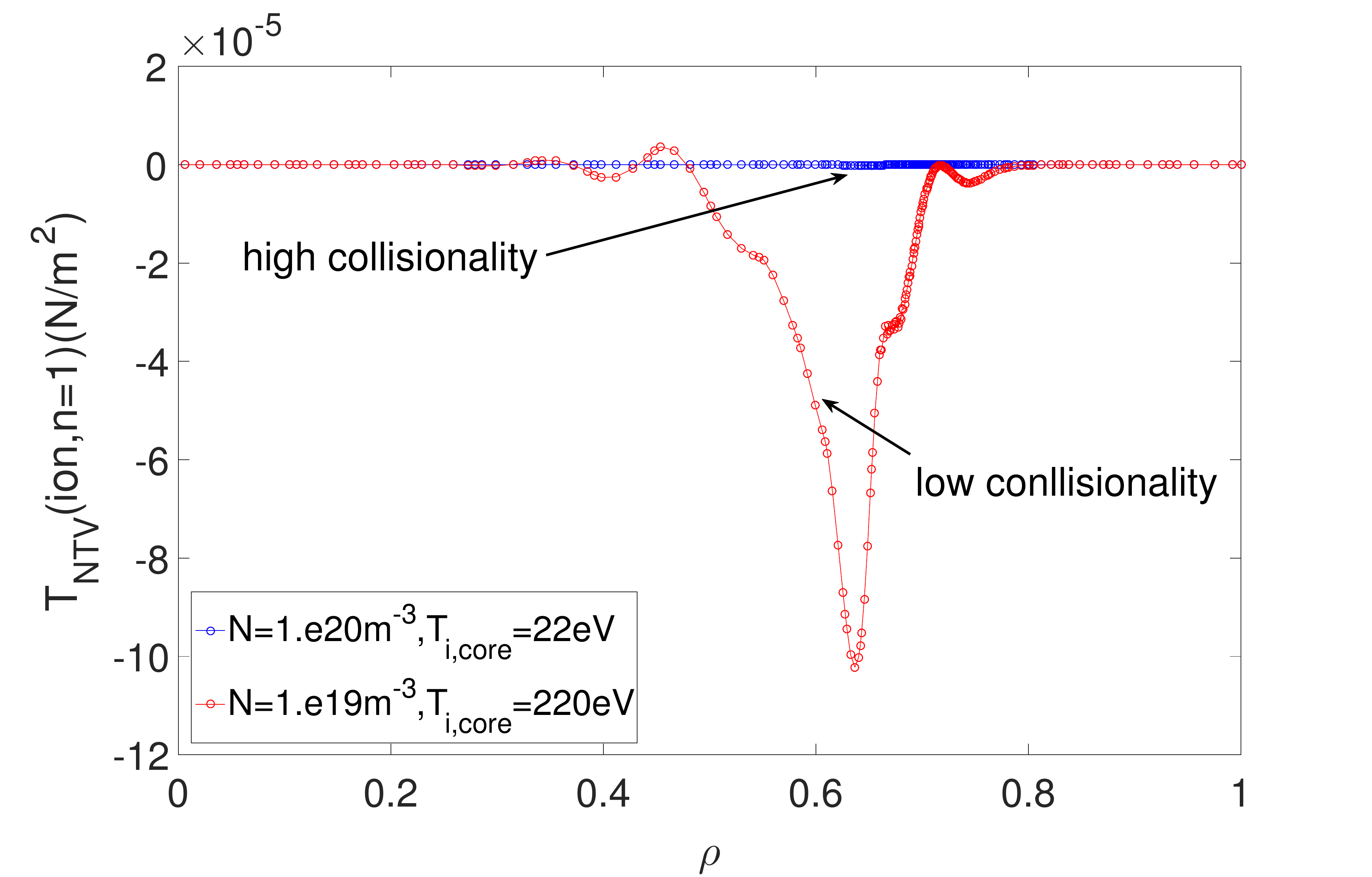}
  \put(-290,180){\textbf{(a)}}
  \vfill
  \includegraphics[width=0.6\textwidth,height=0.29\textheight]{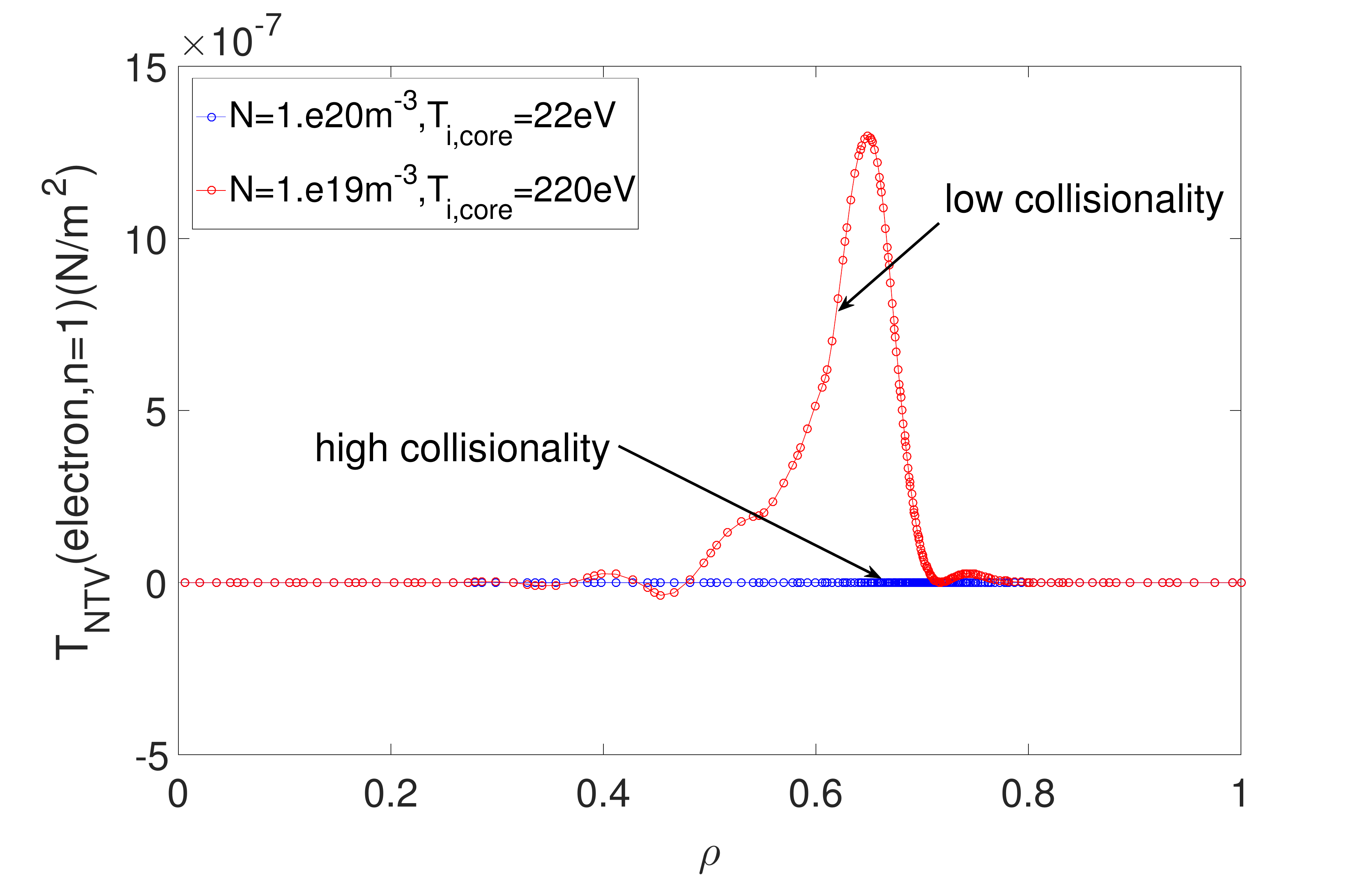}
  \put(-290,180){\textbf{(b)}}
  \caption{Radial profiles as functions of minor radius $\rho$ for (a) ion $T_{NTV}$ and (b) electron $T_{NTV}$ induced by a fixed $n=1$ component of plasma response for equilibria with high (blue lines) and low (red lines) collisionalities, respectively. Here $\beta=0.088\%$ in both high and low collisionality cases.}
  \label{fig_Tntv_col}
\end{figure}

NTV torque also strongly depends on the pedestal collisionality. To study such a dependence, we investigate two cases with different collisionalities in Fig.~\ref{fig_Tntv_col}. In both cases, $\beta=0.088\%$, but the ratios of density to temperature are different, which results in different collisionalities. It shows that with decreasing collisionality, the NTV torque magnitude is substantially increased (shown in red lines) in comparison to the high collisionality case (shown in blue lines). Future advanced tokamaks would typically operate in low collisionality regimes, where the NTV torque is excepted to play a more significant role.

\subsection{Effects of RMP amplitude on NTV torque}

\begin{figure}[ht]
  \includegraphics[width=0.6\textwidth,height=0.28\textheight]{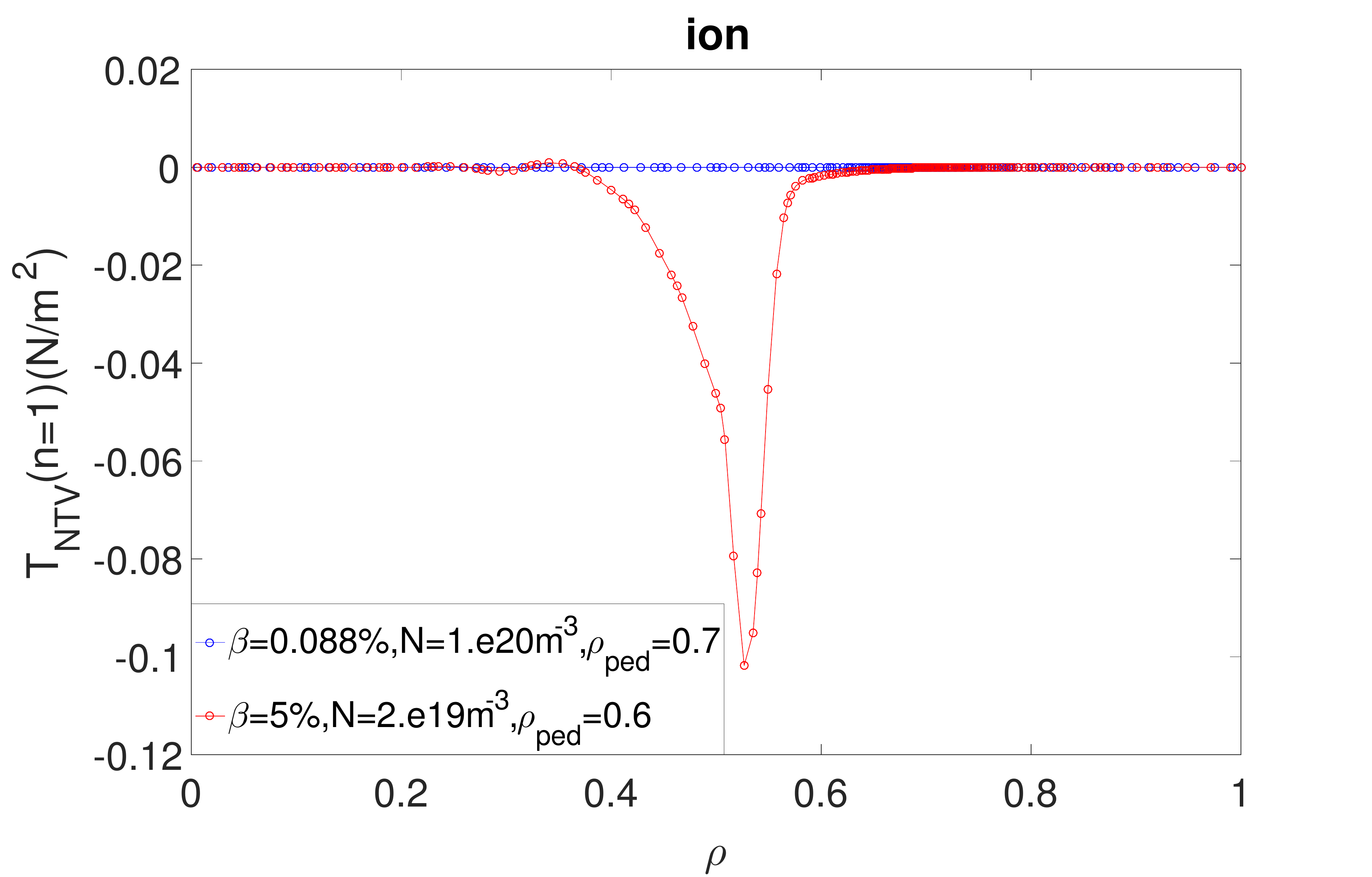}
  \put(-290,180){\textbf{(a)}}
  \vfill
  \includegraphics[width=0.6\textwidth,height=0.28\textheight]{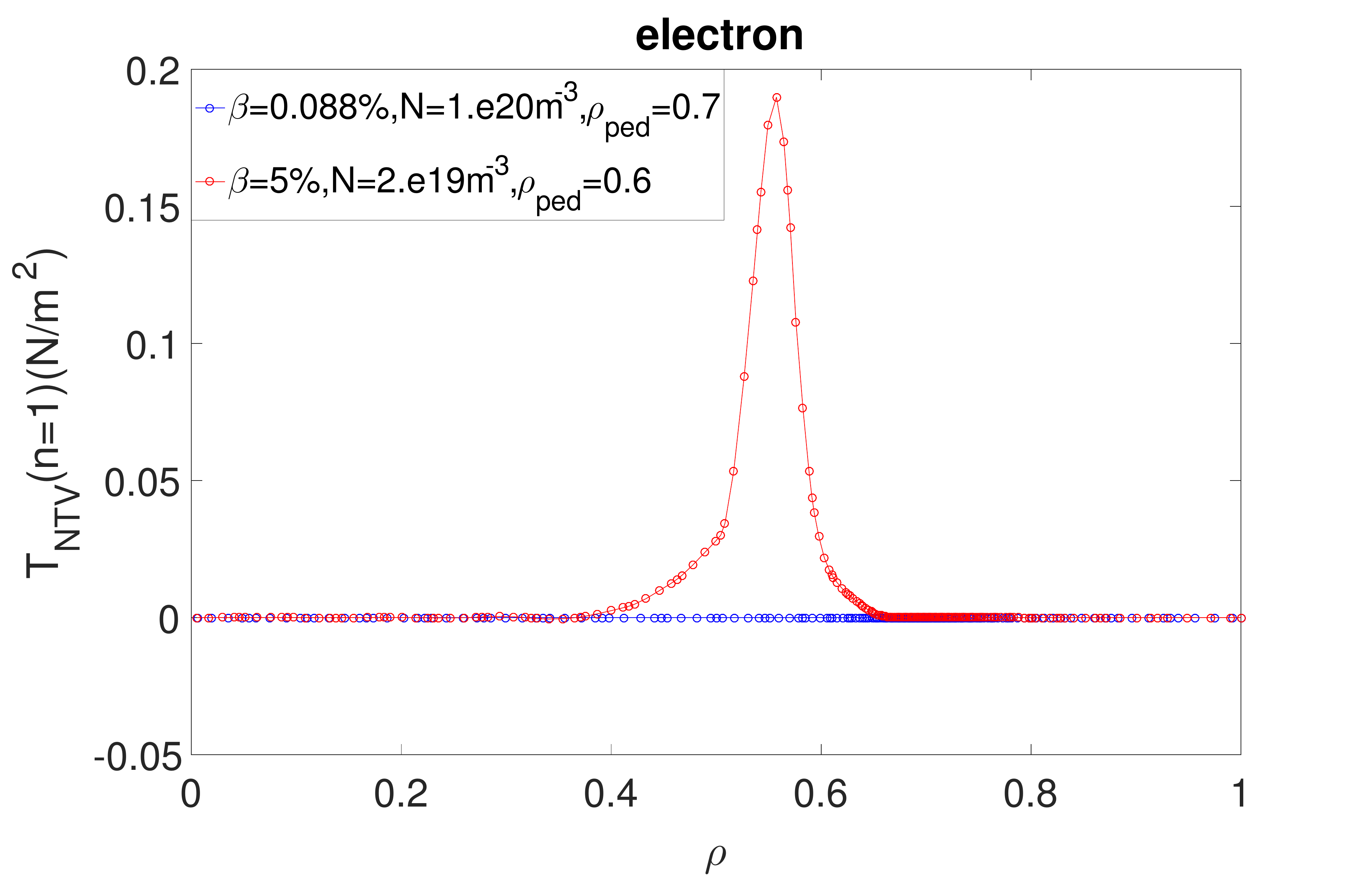}
  \put(-290,180){\textbf{(b)}}
  \vfill
  \includegraphics[width=0.6\textwidth,height=0.28\textheight]{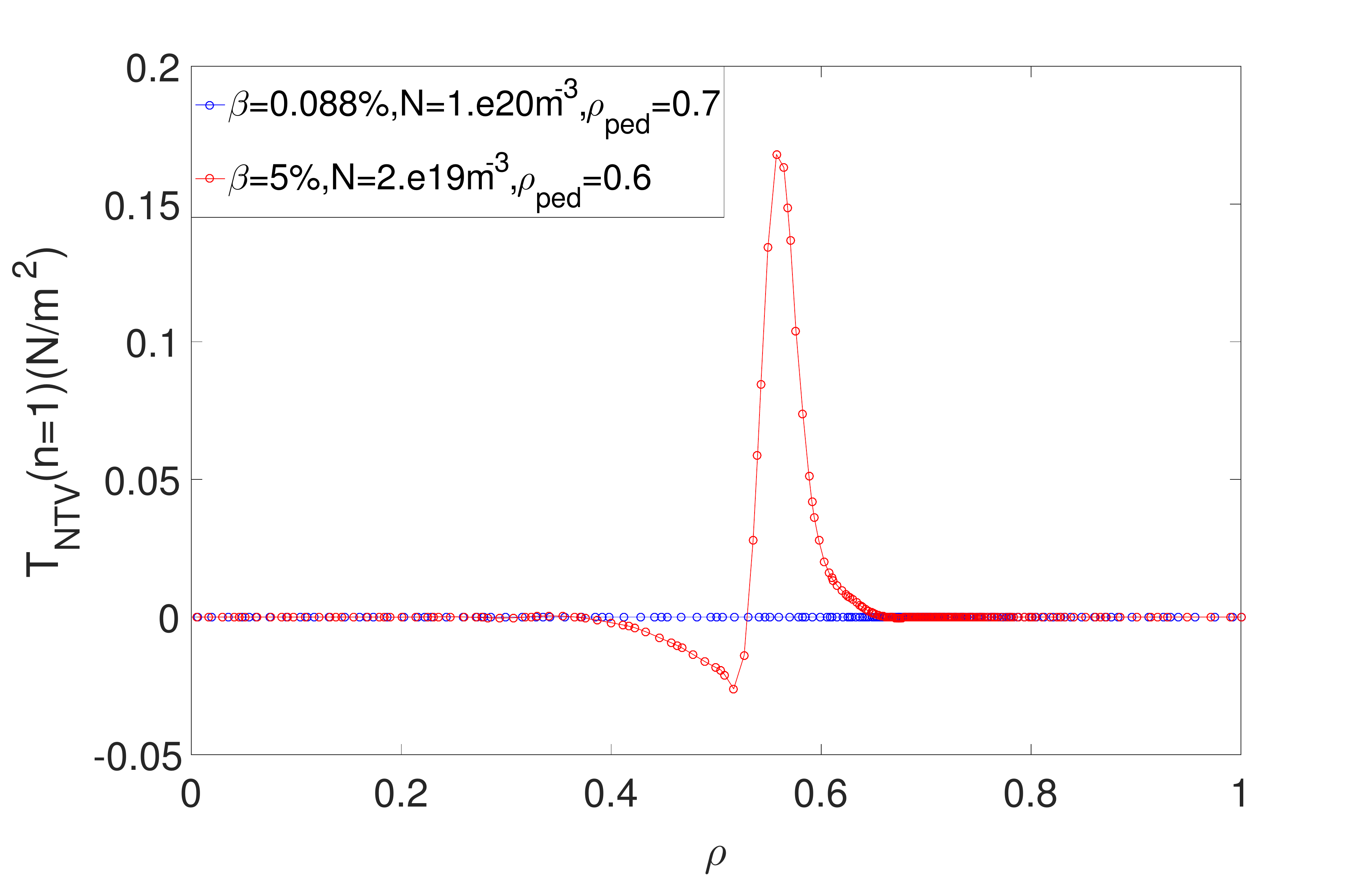}
  \put(-290,180){\textbf{(c)}}
  \caption{Radial profiles as functions of minor radius $\rho$ for: (a) ion $T_{NTV}$; (b) electron $T_{NTV}$; (c) total $T_{NTV}$ induced by a fixed $n=1$ component of plasma response for different equilibrium regimes. Blue lines indicate the results of baseline case.}
  \label{fig_Tntv_NBI}
\end{figure}

Based on the above findings on the pedestal effects on NTV torque, we can search for the equilibrium parameter regimes where the magnitude of NTV torque can be maximal for a given fixed plasma response. One such case is found, where the equilibrium parameters are $\beta=5\%$, $T_{i,core}\approx 6 keV$, $N=2\times 10^{19} m^{-3}$, $\rho_{ped}=0.6$. The NTV torque profiles from this equilibrium, which are plotted as red lines in Fig.~\ref{fig_Tntv_NBI}, are strongly enhanced in comparison to the baseline case. The maximum ion NTV torque magnitude can reach the order of $0.1N/m^2$ (Fig.~\ref{fig_Tntv_NBI}a). Similarly, electron NTV torque is also enhanced in the new equilibrium case (Fig.~\ref{fig_Tntv_NBI}b). However, because the ion NTV torque and electron NTV torque are in opposite directions, and the peak position of ion NTV torque is slightly off from that of electron NTV torque, the ion NTV torque is partially canceled by the electron NTV torque in the total NTV torque profile, which is dominated by the electron contribution (Fig.~\ref{fig_Tntv_NBI}c).

\begin{figure}[ht]
  \includegraphics[width=0.5\textwidth,height=0.27\textheight]{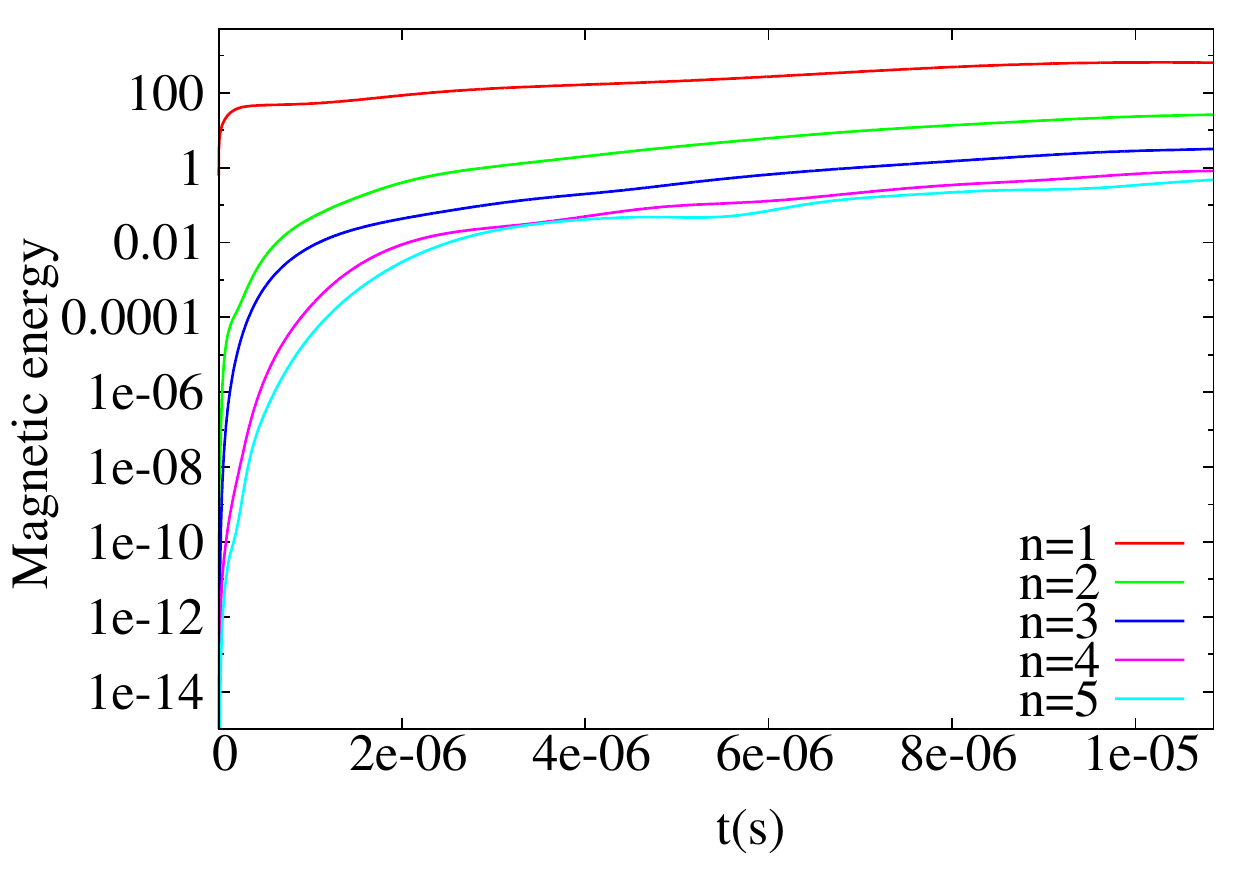}
  \put(-220,180){\textbf{(a)}}
  \hspace{0.65cm}\hfill
  \includegraphics[width=0.46\textwidth,height=0.27\textheight]{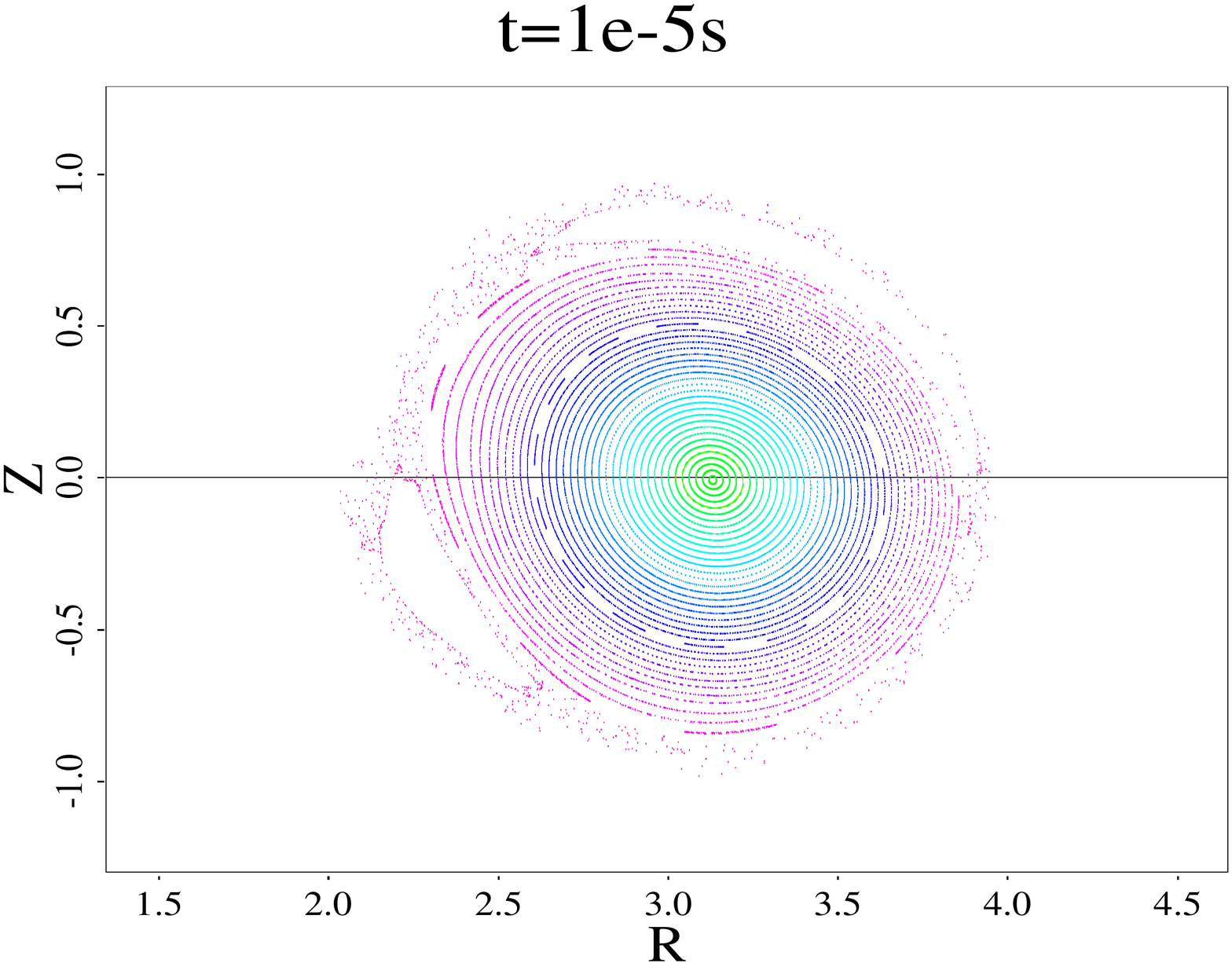}
  \put(-220,180){\textbf{(b)}}
  \vfill
  \includegraphics[width=0.55\textwidth,height=0.27\textheight]{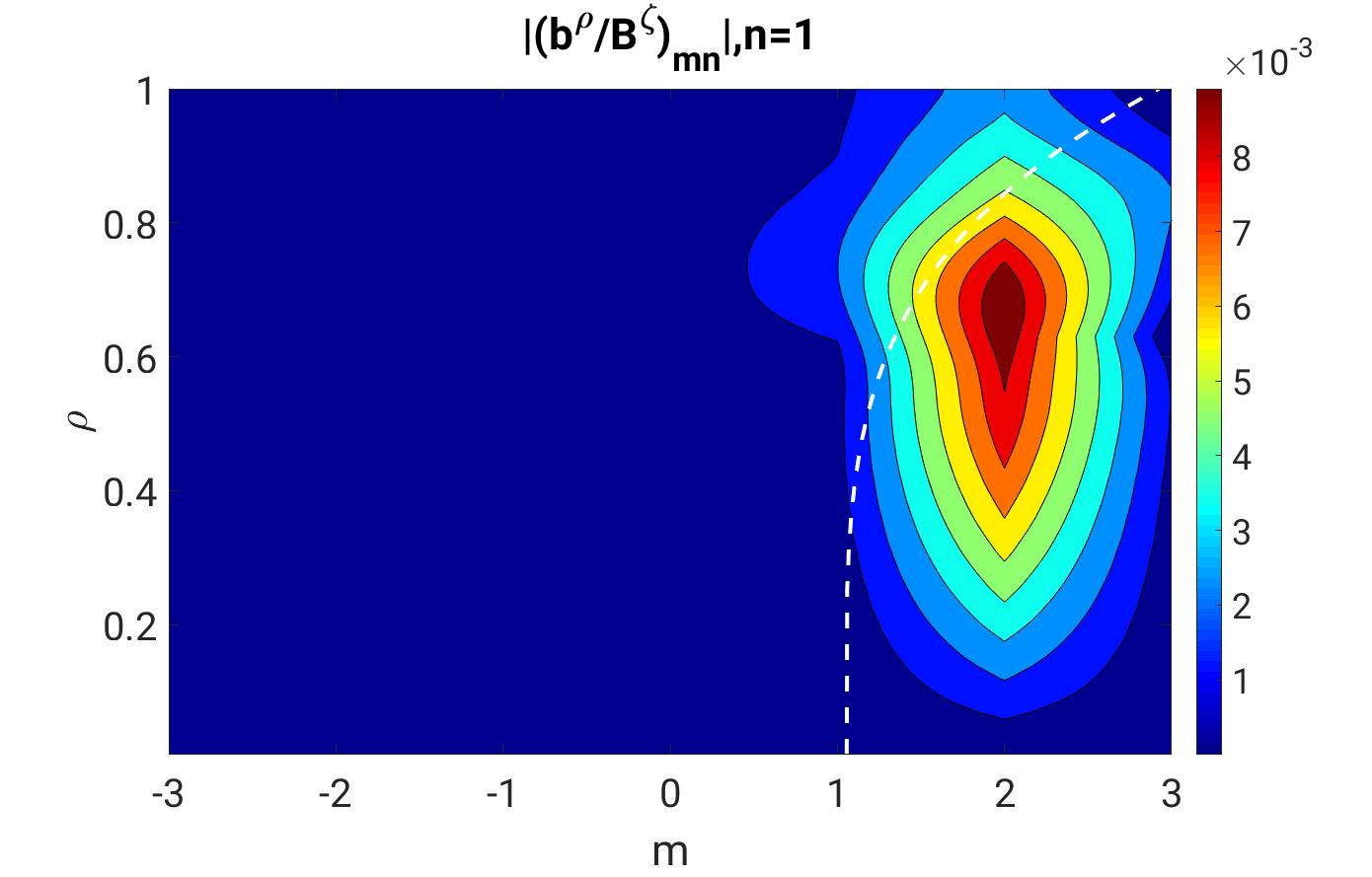}
  \put(-250,170){\textbf{(c)}}
  \includegraphics[width=0.55\textwidth,height=0.27\textheight]{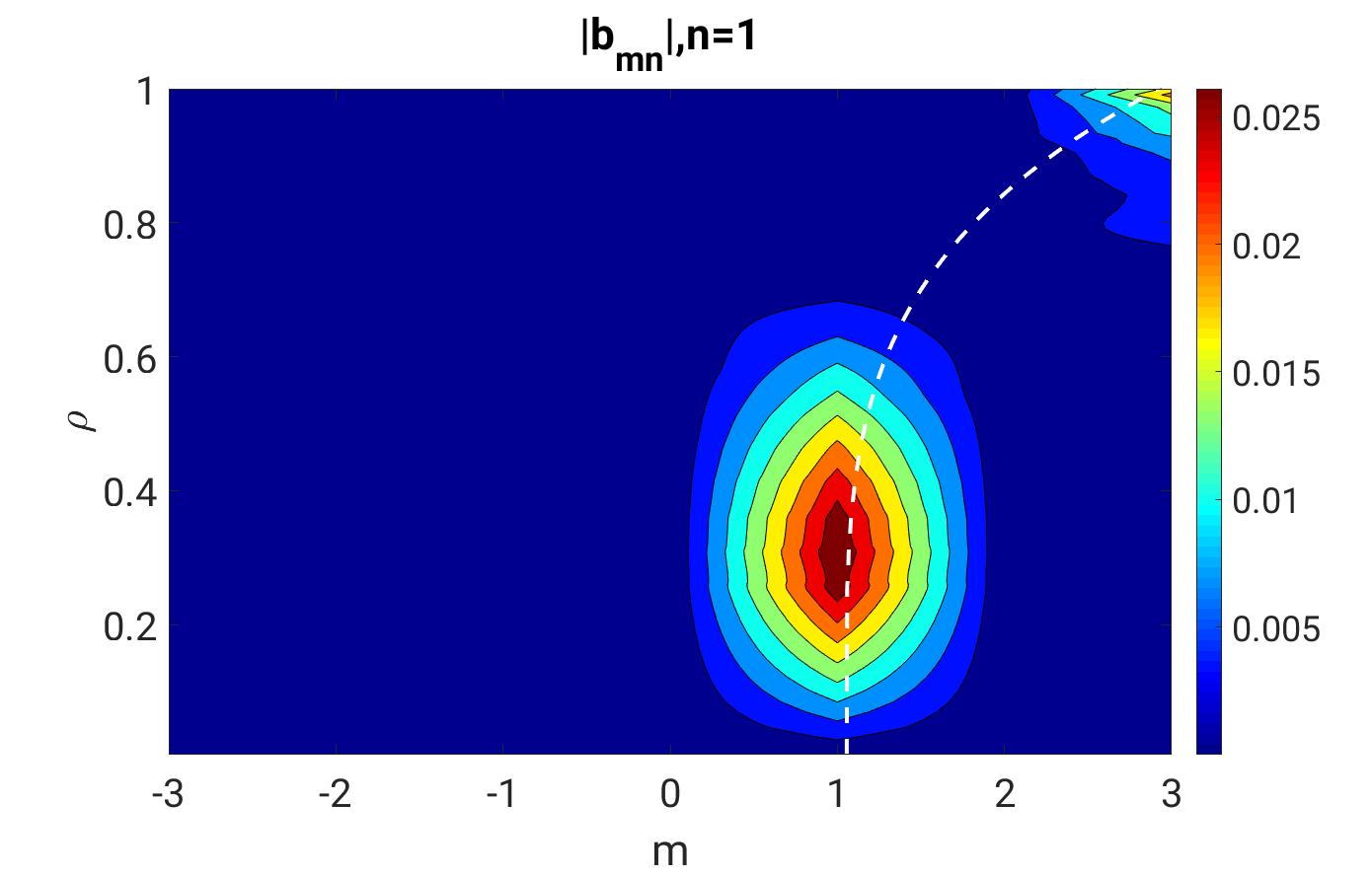}
  \put(-250,170){\textbf{(d)}}
  \caption{(a) Time evolution of perturbed magnetic energies of different toroidal components. (b) Poincare plot at the saturation phase ($t=10^{-5}s$). Contours of (c) radial component of plasma response $|(b^{\rho}/B^{\zeta})_{mn}|$ and (d) perturbed magnetic field strength $|b_{mn}|$ for the $n=1$ component. Here RMP amplitude is $10^{-3}T$. }
  \label{fig_NBI_res}
\end{figure}

In previous calculations, the plasma response is fixed and the RMP amplitude is $10^{-4}T$. However, in real experiments, RMP amplitude can be larger and the corresponding plasma response can be different for different equilibrium profiles. To evaluate such a scenario, we consider the following case: $\beta=1.5\%$, $N=10^{19}m^{-3}$, $T_{i,core}\approx 3.7keV$, $\rho_{ped}=0.65$, the RMP amplitude is $10^{-3}T$, $S=6.5\times 10^7$, and $Pr_m=10$. The magnetic energy evolution (Fig.~\ref{fig_NBI_res}a) shows the dominance of $n=1$ component of plasma response, and the saturated Poincare plot (Fig.~\ref{fig_NBI_res}b) shows the larger $(2,1)$ island and stochastic field lines in the edge region. The contour plots of $|(b^\rho/B^\zeta)_{mn}|$ and $|b_{mn}|$ for $n=1$ component of plasma response (Figs.~\ref{fig_NBI_res}c-d) are similar with those in Fig.~\ref{fig_base_res}, except that the amplitude of the spectrum becomes larger than that in the baseline case. Thus NTV torque is still mainly induced by $(m,n)=(1,1)$ component, which is generated by the distortion of flux surfaces near $q=1$ surface. The resulting NTV torque profile is peaked around $q=1$ surface, which is well away from the $(2,1)$ island and edge stochastic field line region.

\begin{figure}[ht]
  \includegraphics[width=0.6\textwidth,height=0.28\textheight]{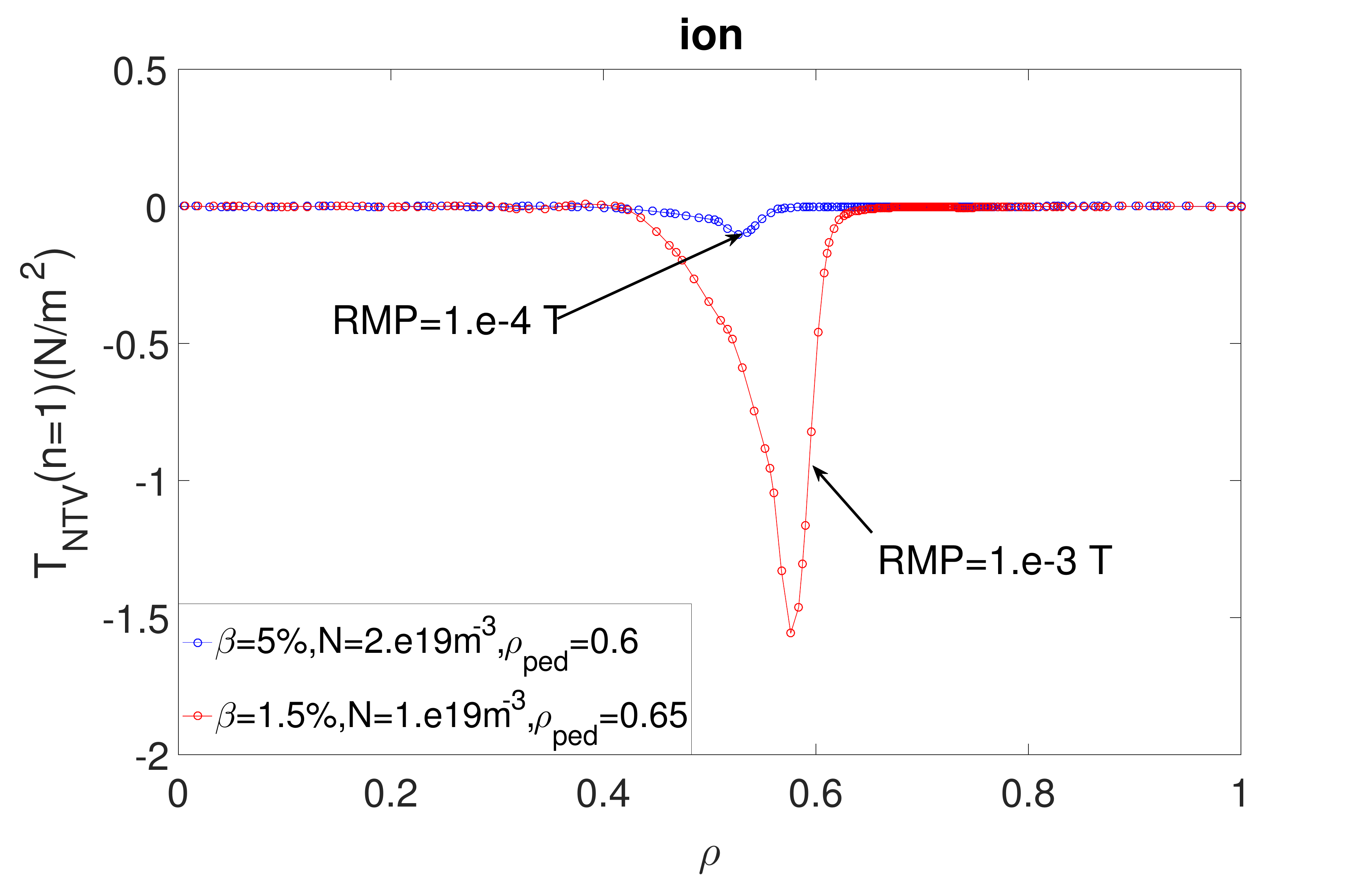}
  \put(-290,180){\textbf{(a)}}
  \vfill
  \includegraphics[width=0.6\textwidth,height=0.28\textheight]{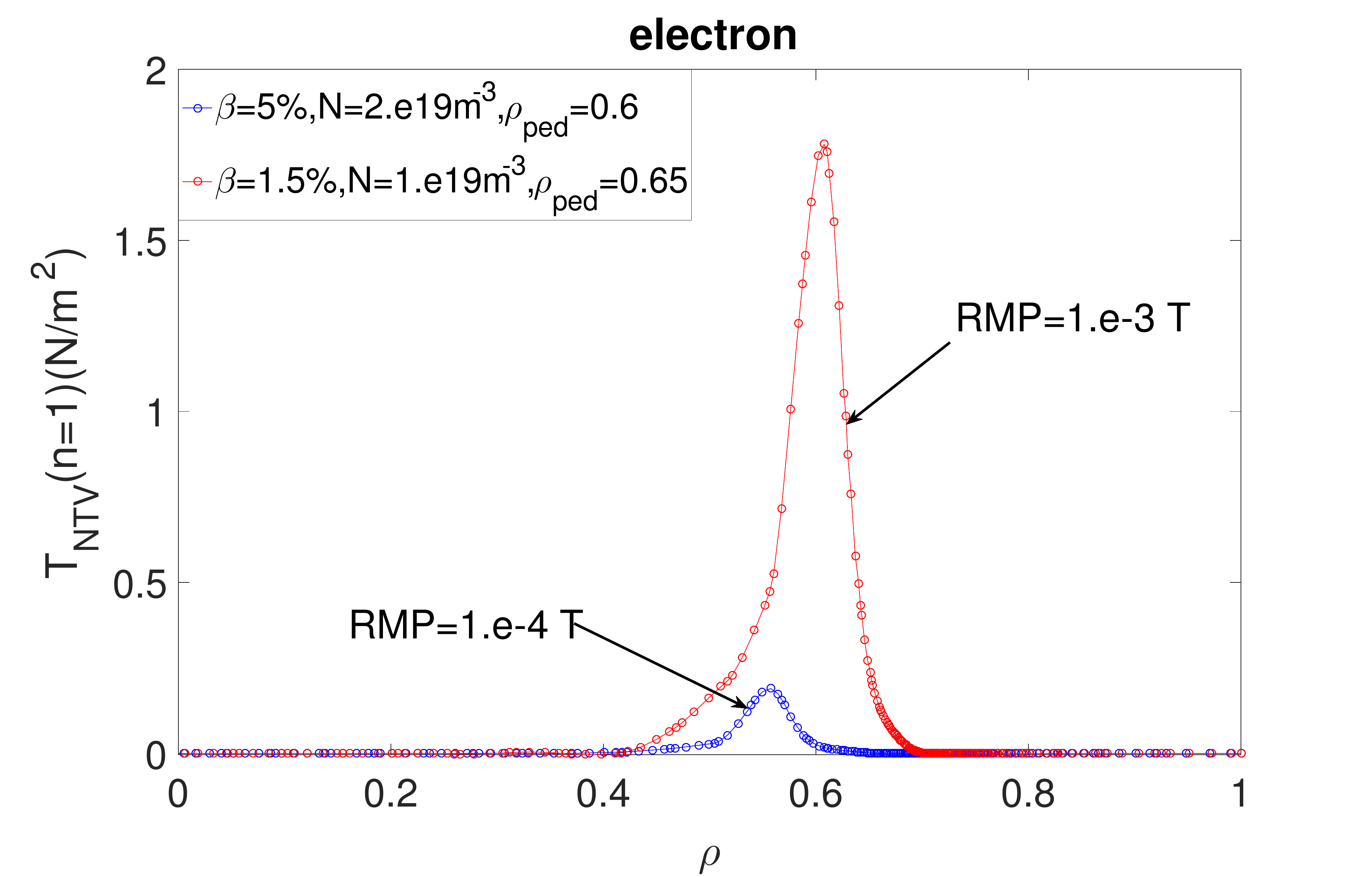}
  \put(-290,180){\textbf{(b)}}
  \vfill
  \includegraphics[width=0.6\textwidth,height=0.28\textheight]{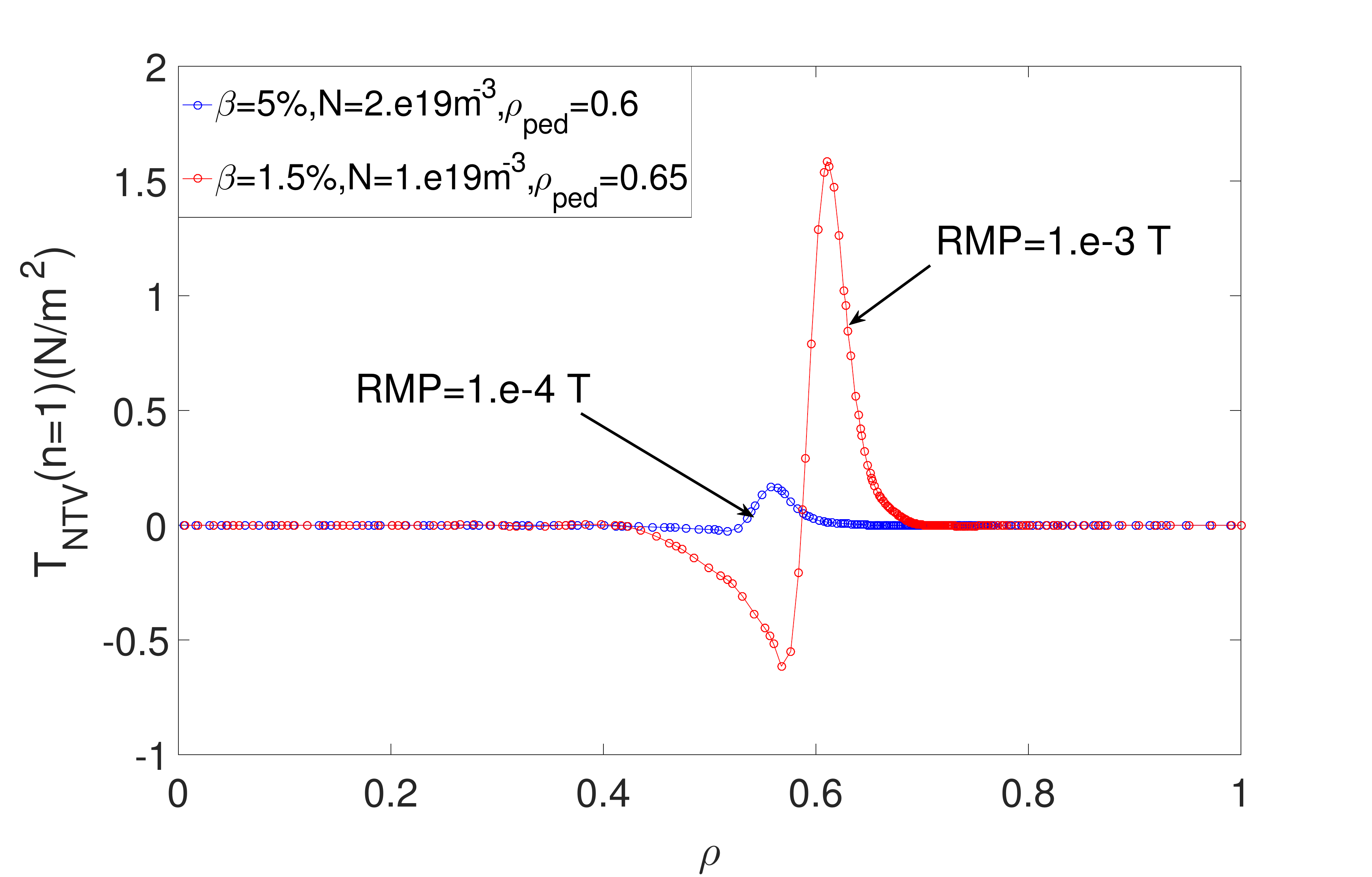}
  \put(-290,180){\textbf{(c)}}
  \caption{Radial profiles as functions of minor radius $\rho$ for: (a) ion $T_{NTV}$; (b) electron $T_{NTV}$; (c) total $T_{NTV}$ induced by the $n=1$ component of plasma response for different RMP amplitudes and equilibrium regimes. Blue lines (RMP amplitude is $10^{-4}T$) represent the same cases as the red lines in Fig.~\ref{fig_Tntv_NBI}.}
  \label{fig_Tntv_NBI6}
\end{figure}

The maximum values of the calculated ion and electron NTV torques are both of the order of $1N/m^2$ (Figs.~\ref{fig_Tntv_NBI6}a-b). As a consequence of the opposite directions of ion and electron NTV torques, there are two peaks in the total NTV torque profile, where the maximum value is $1.5 N/m^2$ (Fig.~\ref{fig_Tntv_NBI6}c). This magnitude of NTV torque is comparable to other common sources of momentum in experiments. For example, neutral beam injection (NBI) can  induce a toroidal torque $\sim 1N/m^2$ \cite{DIII-D_NBI}. In comparison to the case with $10^{-4}T$ RMP amplitude, it's clear that although plasma $\beta$ is moderate, the NTV torque can still reach a larger value with the $10^{-3}T$ RMP amplitude. These results highlight the role of RMP amplitude, which suggests another effective means to enhance NTV torque besides optimizing the pedestal profiles.

We note that in both Figs.~\ref{fig_Tntv_NBI} and \ref{fig_Tntv_NBI6}, the amplitudes of electron NTV torque are comparable, or even dominant in comparison to the ion NTV torque. Generally speaking, electron NTV torque can become important in low collisionality regimes \cite{Sun11_nf}. Because electron NTV torque can be in opposite direction to ion NTV torque, once their amplitudes are comparable, the total NTV torque profile could be quite different from the ion/electron NTV torque profiles alone.

\section{Summary and discussions}
\label{sec:sum}

In summary, we have developed an initial coupling between the NIMROD and the NTVTOK codes to calculate NTV torque induced by realistic NAMPs in tokamaks. In this work, we have evaluated the NTV torque in edge pedestal induced by plasma response to RMP in particular. We find that NTV torque profile is determined by the combined effects of plasma response and temperature (pressure) profiles. NTV torque is found to increase when the peak positions of plasma response and pressure pedestal gradient become closer spatially, as also indicated in the approximated formula for NTV torque profile in Eq.~(\ref{eq_Tntv_approx}). The amplitude of NTV torque scales strongly with pedestal $\beta$, which suggests the importance of NTV torque in the high-$\beta$ H-mode regimes. For fixed $\beta$, decreasing collisionality can also enhance the magnitude of NTV torque in the $1/\nu$ regime. In addition to equilibrium profiles, changing RMP amplitude can also effectively influence NTV torque. Based on these findings, we have identified the tokamak operation regimes where NTV torque can be comparable in magnitude with the torque generated by neutral beam injection. Thus, NTV torque could potentially play a significant role in tokamak edge dynamics.

In most of the current analyses, plasma response to RMP is fixed for the sake of clarity. However, in fact, the plasma response will also change along with the equilibrium parameters and profiles. The indirect effects of equilibrium on NTV torque through the corresponding plasma response should be investigated next, which is beyond the scope of current report. Furthermore, a more self-consistent calculation of plasma response to RMP as well as the resulting NTV torque should be based on a more complete MHD model, where the NTV torque itself is also included in the momentum evolution equation. We plan on exploring such a possibility of development in future work.

\begin{acknowledgments}
This work was supported by National Magnetic Confinement Fusion Science Program of China under Grant Nos. 2014GB124002 and 2015GB101004. U.S. Department of Energy Grant Nos. DE-FG02-86ER53218 and DE-FC02-08ER54975, and the 100 Talent Program of the Chinese Academy of Sciences. 
This research used the computing resources from the Supercomputing Center of University of Science and Technology of China, and the National Energy Research Scientific Computing Center, a DOE Office of Science User Facility supported by the Office of Science of the U.S. Department of Energy under Contract No. DE-AC02-05CH11231.
\end{acknowledgments}

\bibliographystyle{apsrev}
\bibliography{yan}

\begin{thebibliography}{43}
\expandafter\ifx\csname natexlab\endcsname\relax\def\natexlab#1{#1}\fi
\expandafter\ifx\csname bibnamefont\endcsname\relax
  \def\bibnamefont#1{#1}\fi
\expandafter\ifx\csname bibfnamefont\endcsname\relax
  \def\bibfnamefont#1{#1}\fi
\expandafter\ifx\csname citenamefont\endcsname\relax
  \def\citenamefont#1{#1}\fi
\expandafter\ifx\csname url\endcsname\relax
  \def\url#1{\texttt{#1}}\fi
\expandafter\ifx\csname urlprefix\endcsname\relax\def\urlprefix{URL }\fi
\providecommand{\bibinfo}[2]{#2}
\providecommand{\eprint}[2][]{\url{#2}}

\bibitem[{\citenamefont{Terry}(2000)}]{terry00a}
\bibinfo{author}{\bibfnamefont{P.~W.} \bibnamefont{Terry}},
  \bibinfo{journal}{Reviews of Modern Physics} \textbf{\bibinfo{volume}{72}},
  \bibinfo{pages}{109} (\bibinfo{year}{2000}).

\bibitem[{\citenamefont{Itoh and Itoh}(1988)}]{itoh88a}
\bibinfo{author}{\bibfnamefont{S.-I.} \bibnamefont{Itoh}} \bibnamefont{and}
  \bibinfo{author}{\bibfnamefont{K.}~\bibnamefont{Itoh}},
  \bibinfo{journal}{Physical Review Letters} \textbf{\bibinfo{volume}{60}},
  \bibinfo{pages}{2276} (\bibinfo{year}{1988}).

\bibitem[{\citenamefont{Shaing and Crume}(1989)}]{shaing89a}
\bibinfo{author}{\bibfnamefont{K.~C.} \bibnamefont{Shaing}} \bibnamefont{and}
  \bibinfo{author}{\bibfnamefont{E.~C.} \bibnamefont{Crume}},
  \bibinfo{journal}{Physical Review Letters} \textbf{\bibinfo{volume}{63}},
  \bibinfo{pages}{2369} (\bibinfo{year}{1989}).

\bibitem[{\citenamefont{Shaing et~al.}(1990)\citenamefont{Shaing, Crume, and
  Houlberg}}]{shaing90a}
\bibinfo{author}{\bibfnamefont{K.~C.} \bibnamefont{Shaing}},
  \bibinfo{author}{\bibfnamefont{E.~C.} \bibnamefont{Crume}}, \bibnamefont{and}
  \bibinfo{author}{\bibfnamefont{W.~A.} \bibnamefont{Houlberg}},
  \bibinfo{journal}{Physics of Fluids B} \textbf{\bibinfo{volume}{2}},
  \bibinfo{pages}{1492} (\bibinfo{year}{1990}).

\bibitem[{\citenamefont{Oyama et~al.}(2005)\citenamefont{Oyama, Sakamoto,
  Isayama, Takechi, Gohil, Lao, Snyder, Fujita, Ide, Kamada et~al.}}]{Oyama}
\bibinfo{author}{\bibfnamefont{N.}~\bibnamefont{Oyama}},
  \bibinfo{author}{\bibfnamefont{Y.}~\bibnamefont{Sakamoto}},
  \bibinfo{author}{\bibfnamefont{A.}~\bibnamefont{Isayama}},
  \bibinfo{author}{\bibfnamefont{M.}~\bibnamefont{Takechi}},
  \bibinfo{author}{\bibfnamefont{P.}~\bibnamefont{Gohil}},
  \bibinfo{author}{\bibfnamefont{L.}~\bibnamefont{Lao}},
  \bibinfo{author}{\bibfnamefont{P.~B.} \bibnamefont{Snyder}},
  \bibinfo{author}{\bibfnamefont{T.}~\bibnamefont{Fujita}},
  \bibinfo{author}{\bibfnamefont{S.}~\bibnamefont{Ide}},
  \bibinfo{author}{\bibfnamefont{Y.}~\bibnamefont{Kamada}},
  \bibnamefont{et~al.}, \bibinfo{journal}{Nuclear Fusion}
  \textbf{\bibinfo{volume}{45}}, \bibinfo{pages}{871} (\bibinfo{year}{2005}).

\bibitem[{\citenamefont{Burrell et~al.}(2009)\citenamefont{Burrell, Osborne,
  Snyder, West, Fenstermacher, Groebner, Gohil, Leonard, and
  Solomon}}]{Burrell_1}
\bibinfo{author}{\bibfnamefont{K.~H.} \bibnamefont{Burrell}},
  \bibinfo{author}{\bibfnamefont{T.~H.} \bibnamefont{Osborne}},
  \bibinfo{author}{\bibfnamefont{P.~B.} \bibnamefont{Snyder}},
  \bibinfo{author}{\bibfnamefont{W.~P.} \bibnamefont{West}},
  \bibinfo{author}{\bibfnamefont{M.~E.} \bibnamefont{Fenstermacher}},
  \bibinfo{author}{\bibfnamefont{R.~J.} \bibnamefont{Groebner}},
  \bibinfo{author}{\bibfnamefont{P.}~\bibnamefont{Gohil}},
  \bibinfo{author}{\bibfnamefont{A.~W.} \bibnamefont{Leonard}},
  \bibnamefont{and} \bibinfo{author}{\bibfnamefont{W.~M.}
  \bibnamefont{Solomon}}, \bibinfo{journal}{Nuclear Fusion}
  \textbf{\bibinfo{volume}{49}}, \bibinfo{pages}{085024}
  (\bibinfo{year}{2009}).

\bibitem[{\citenamefont{Evans et~al.}(2004)\citenamefont{Evans, Moyer, Thomas,
  Watkins, Osborne, Boedo, Doyle, Fenstermacher, Finken, Groebner
  et~al.}}]{Evans}
\bibinfo{author}{\bibfnamefont{T.~E.} \bibnamefont{Evans}},
  \bibinfo{author}{\bibfnamefont{R.~A.} \bibnamefont{Moyer}},
  \bibinfo{author}{\bibfnamefont{P.~R.} \bibnamefont{Thomas}},
  \bibinfo{author}{\bibfnamefont{J.~G.} \bibnamefont{Watkins}},
  \bibinfo{author}{\bibfnamefont{T.~H.} \bibnamefont{Osborne}},
  \bibinfo{author}{\bibfnamefont{J.~A.} \bibnamefont{Boedo}},
  \bibinfo{author}{\bibfnamefont{E.~J.} \bibnamefont{Doyle}},
  \bibinfo{author}{\bibfnamefont{M.~E.} \bibnamefont{Fenstermacher}},
  \bibinfo{author}{\bibfnamefont{K.~H.} \bibnamefont{Finken}},
  \bibinfo{author}{\bibfnamefont{R.~J.} \bibnamefont{Groebner}},
  \bibnamefont{et~al.}, \bibinfo{journal}{Physical Review Letters}
  \textbf{\bibinfo{volume}{92}}, \bibinfo{pages}{235003}
  (\bibinfo{year}{2004}).

\bibitem[{\citenamefont{Burrell et~al.}(2005)\citenamefont{Burrell, Evans,
  Doyle, Fenstermacher, Groebner, Leonard, Moyer, Osborne, Schaffer, Snyder
  et~al.}}]{Burrell_2}
\bibinfo{author}{\bibfnamefont{K.~H.} \bibnamefont{Burrell}},
  \bibinfo{author}{\bibfnamefont{T.~E.} \bibnamefont{Evans}},
  \bibinfo{author}{\bibfnamefont{E.~J.} \bibnamefont{Doyle}},
  \bibinfo{author}{\bibfnamefont{M.~E.} \bibnamefont{Fenstermacher}},
  \bibinfo{author}{\bibfnamefont{R.~J.} \bibnamefont{Groebner}},
  \bibinfo{author}{\bibfnamefont{A.~W.} \bibnamefont{Leonard}},
  \bibinfo{author}{\bibfnamefont{R.~A.} \bibnamefont{Moyer}},
  \bibinfo{author}{\bibfnamefont{T.~H.} \bibnamefont{Osborne}},
  \bibinfo{author}{\bibfnamefont{M.~J.} \bibnamefont{Schaffer}},
  \bibinfo{author}{\bibfnamefont{P.~B.} \bibnamefont{Snyder}},
  \bibnamefont{et~al.}, \bibinfo{journal}{Plasma Physics and Controlled Fusion}
  \textbf{\bibinfo{volume}{47}}, \bibinfo{pages}{B37} (\bibinfo{year}{2005}).

\bibitem[{\citenamefont{Liang et~al.}(2007)\citenamefont{Liang, Koslowski,
  Thomas, Nardon, Alper, Andrew, Andrew, Arnoux, Baranov, B\'ecoulet
  et~al.}}]{Liang}
\bibinfo{author}{\bibfnamefont{Y.}~\bibnamefont{Liang}},
  \bibinfo{author}{\bibfnamefont{H.~R.} \bibnamefont{Koslowski}},
  \bibinfo{author}{\bibfnamefont{P.~R.} \bibnamefont{Thomas}},
  \bibinfo{author}{\bibfnamefont{E.}~\bibnamefont{Nardon}},
  \bibinfo{author}{\bibfnamefont{B.}~\bibnamefont{Alper}},
  \bibinfo{author}{\bibfnamefont{P.}~\bibnamefont{Andrew}},
  \bibinfo{author}{\bibfnamefont{Y.}~\bibnamefont{Andrew}},
  \bibinfo{author}{\bibfnamefont{G.}~\bibnamefont{Arnoux}},
  \bibinfo{author}{\bibfnamefont{Y.}~\bibnamefont{Baranov}},
  \bibinfo{author}{\bibfnamefont{M.}~\bibnamefont{B\'ecoulet}},
  \bibnamefont{et~al.}, \bibinfo{journal}{Physical Review Letters}
  \textbf{\bibinfo{volume}{98}}, \bibinfo{pages}{265004}
  (\bibinfo{year}{2007}).

\bibitem[{\citenamefont{Evans}(2015)}]{Evans_2015_PPCF}
\bibinfo{author}{\bibfnamefont{T.~E.} \bibnamefont{Evans}},
  \bibinfo{journal}{Plasma Physics and Controlled Fusion}
  \textbf{\bibinfo{volume}{57}}, \bibinfo{pages}{123001}
  (\bibinfo{year}{2015}).

\bibitem[{\citenamefont{Nazikian et~al.}(2015)\citenamefont{Nazikian,
  Paz-Soldan, Callen, deGrassie, Eldon, Evans, Ferraro, Grierson, Groebner,
  Haskey et~al.}}]{Nazikian_2015_PRL}
\bibinfo{author}{\bibfnamefont{R.}~\bibnamefont{Nazikian}},
  \bibinfo{author}{\bibfnamefont{C.}~\bibnamefont{Paz-Soldan}},
  \bibinfo{author}{\bibfnamefont{J.~D.} \bibnamefont{Callen}},
  \bibinfo{author}{\bibfnamefont{J.~S.} \bibnamefont{deGrassie}},
  \bibinfo{author}{\bibfnamefont{D.}~\bibnamefont{Eldon}},
  \bibinfo{author}{\bibfnamefont{T.~E.} \bibnamefont{Evans}},
  \bibinfo{author}{\bibfnamefont{N.~M.} \bibnamefont{Ferraro}},
  \bibinfo{author}{\bibfnamefont{B.~A.} \bibnamefont{Grierson}},
  \bibinfo{author}{\bibfnamefont{R.~J.} \bibnamefont{Groebner}},
  \bibinfo{author}{\bibfnamefont{S.~R.} \bibnamefont{Haskey}},
  \bibnamefont{et~al.}, \bibinfo{journal}{Physical Review Letters}
  \textbf{\bibinfo{volume}{114}}, \bibinfo{pages}{105002}
  (\bibinfo{year}{2015}).

\bibitem[{\citenamefont{Callen et~al.}()\citenamefont{Callen, Nazikian,
  Paz-Soldan, Ferraro, Beidler, Hegna, and La-Haye}}]{Callen_CPTC}
\bibinfo{author}{\bibfnamefont{J.~D.} \bibnamefont{Callen}},
  \bibinfo{author}{\bibfnamefont{R.}~\bibnamefont{Nazikian}},
  \bibinfo{author}{\bibfnamefont{C.}~\bibnamefont{Paz-Soldan}},
  \bibinfo{author}{\bibfnamefont{N.~M.} \bibnamefont{Ferraro}},
  \bibinfo{author}{\bibfnamefont{M.~T.} \bibnamefont{Beidler}},
  \bibinfo{author}{\bibfnamefont{C.~C.} \bibnamefont{Hegna}}, \bibnamefont{and}
  \bibinfo{author}{\bibfnamefont{R.~J.} \bibnamefont{La-Haye}},
  \bibinfo{note}{{R}eport No. UW-CPTC 16-4 (to be published)}.

\bibitem[{\citenamefont{Shaing}(2003)}]{Shaing03}
\bibinfo{author}{\bibfnamefont{K.~C.} \bibnamefont{Shaing}},
  \bibinfo{journal}{Physics of Plasmas} \textbf{\bibinfo{volume}{10}},
  \bibinfo{pages}{1443} (\bibinfo{year}{2003}).

\bibitem[{\citenamefont{Park et~al.}(2009{\natexlab{a}})\citenamefont{Park,
  Boozer, and Menard}}]{Park09}
\bibinfo{author}{\bibfnamefont{J.-K.} \bibnamefont{Park}},
  \bibinfo{author}{\bibfnamefont{A.~H.} \bibnamefont{Boozer}},
  \bibnamefont{and} \bibinfo{author}{\bibfnamefont{J.~E.}
  \bibnamefont{Menard}}, \bibinfo{journal}{Physical Review Letters}
  \textbf{\bibinfo{volume}{102}}, \bibinfo{pages}{065002}
  (\bibinfo{year}{2009}{\natexlab{a}}).

\bibitem[{\citenamefont{Zhu et~al.}(2006)\citenamefont{Zhu, Sabbagh, Bell,
  Bialek, Bell, LeBlanc, Kaye, Levinton, Menard, Shaing et~al.}}]{Zhu}
\bibinfo{author}{\bibfnamefont{W.}~\bibnamefont{Zhu}},
  \bibinfo{author}{\bibfnamefont{S.~A.} \bibnamefont{Sabbagh}},
  \bibinfo{author}{\bibfnamefont{R.~E.} \bibnamefont{Bell}},
  \bibinfo{author}{\bibfnamefont{J.~M.} \bibnamefont{Bialek}},
  \bibinfo{author}{\bibfnamefont{M.~G.} \bibnamefont{Bell}},
  \bibinfo{author}{\bibfnamefont{B.~P.} \bibnamefont{LeBlanc}},
  \bibinfo{author}{\bibfnamefont{S.~M.} \bibnamefont{Kaye}},
  \bibinfo{author}{\bibfnamefont{F.~M.} \bibnamefont{Levinton}},
  \bibinfo{author}{\bibfnamefont{J.~E.} \bibnamefont{Menard}},
  \bibinfo{author}{\bibfnamefont{K.~C.} \bibnamefont{Shaing}},
  \bibnamefont{et~al.}, \bibinfo{journal}{Physical Review Letters}
  \textbf{\bibinfo{volume}{96}}, \bibinfo{pages}{225002}
  (\bibinfo{year}{2006}).

\bibitem[{\citenamefont{Garofalo et~al.}(2008)\citenamefont{Garofalo, Burrell,
  DeBoo, deGrassie, Jackson, Lanctot, Reimerdes, Schaffer, Solomon, and
  Strait}}]{Garofalo}
\bibinfo{author}{\bibfnamefont{A.~M.} \bibnamefont{Garofalo}},
  \bibinfo{author}{\bibfnamefont{K.~H.} \bibnamefont{Burrell}},
  \bibinfo{author}{\bibfnamefont{J.~C.} \bibnamefont{DeBoo}},
  \bibinfo{author}{\bibfnamefont{J.~S.} \bibnamefont{deGrassie}},
  \bibinfo{author}{\bibfnamefont{G.~L.} \bibnamefont{Jackson}},
  \bibinfo{author}{\bibfnamefont{M.}~\bibnamefont{Lanctot}},
  \bibinfo{author}{\bibfnamefont{H.}~\bibnamefont{Reimerdes}},
  \bibinfo{author}{\bibfnamefont{M.~J.} \bibnamefont{Schaffer}},
  \bibinfo{author}{\bibfnamefont{W.~M.} \bibnamefont{Solomon}},
  \bibnamefont{and} \bibinfo{author}{\bibfnamefont{E.~J.}
  \bibnamefont{Strait}}, \bibinfo{journal}{Physical Review Letters}
  \textbf{\bibinfo{volume}{101}}, \bibinfo{pages}{195005}
  (\bibinfo{year}{2008}).

\bibitem[{\citenamefont{Sabbagh et~al.}(2010)\citenamefont{Sabbagh, Berkery,
  Bell, Bialek, Gerhardt, Menard, Betti, Gates, Hu, Katsuro-Hopkins
  et~al.}}]{Sabbagh}
\bibinfo{author}{\bibfnamefont{S.~A.} \bibnamefont{Sabbagh}},
  \bibinfo{author}{\bibfnamefont{J.~W.} \bibnamefont{Berkery}},
  \bibinfo{author}{\bibfnamefont{R.~E.} \bibnamefont{Bell}},
  \bibinfo{author}{\bibfnamefont{J.~M.} \bibnamefont{Bialek}},
  \bibinfo{author}{\bibfnamefont{S.~P.} \bibnamefont{Gerhardt}},
  \bibinfo{author}{\bibfnamefont{J.~E.} \bibnamefont{Menard}},
  \bibinfo{author}{\bibfnamefont{R.}~\bibnamefont{Betti}},
  \bibinfo{author}{\bibfnamefont{D.~A.} \bibnamefont{Gates}},
  \bibinfo{author}{\bibfnamefont{B.}~\bibnamefont{Hu}},
  \bibinfo{author}{\bibfnamefont{O.~N.} \bibnamefont{Katsuro-Hopkins}},
  \bibnamefont{et~al.}, \bibinfo{journal}{Nuclear Fusion}
  \textbf{\bibinfo{volume}{50}}, \bibinfo{pages}{025020}
  (\bibinfo{year}{2010}).

\bibitem[{\citenamefont{Hua et~al.}(2010)\citenamefont{Hua, Chapman, Field,
  Hastie, Pinches, and the MAST~Team}}]{Hua}
\bibinfo{author}{\bibfnamefont{M.-D.} \bibnamefont{Hua}},
  \bibinfo{author}{\bibfnamefont{I.~T.} \bibnamefont{Chapman}},
  \bibinfo{author}{\bibfnamefont{A.~R.} \bibnamefont{Field}},
  \bibinfo{author}{\bibfnamefont{R.~J.} \bibnamefont{Hastie}},
  \bibinfo{author}{\bibfnamefont{S.~D.} \bibnamefont{Pinches}},
  \bibnamefont{and} \bibinfo{author}{\bibnamefont{the MAST~Team}},
  \bibinfo{journal}{Plasma Physics and Controlled Fusion}
  \textbf{\bibinfo{volume}{52}}, \bibinfo{pages}{035009}
  (\bibinfo{year}{2010}).

\bibitem[{\citenamefont{Sun et~al.}(2010{\natexlab{a}})\citenamefont{Sun,
  Liang, Koslowski, Jachmich, Alfier, Asunta, Corrigan, Giroud, Gryaznevich,
  Harting et~al.}}]{Sun_jet}
\bibinfo{author}{\bibfnamefont{Y.}~\bibnamefont{Sun}},
  \bibinfo{author}{\bibfnamefont{Y.}~\bibnamefont{Liang}},
  \bibinfo{author}{\bibfnamefont{H.~R.} \bibnamefont{Koslowski}},
  \bibinfo{author}{\bibfnamefont{S.}~\bibnamefont{Jachmich}},
  \bibinfo{author}{\bibfnamefont{A.}~\bibnamefont{Alfier}},
  \bibinfo{author}{\bibfnamefont{O.}~\bibnamefont{Asunta}},
  \bibinfo{author}{\bibfnamefont{G.}~\bibnamefont{Corrigan}},
  \bibinfo{author}{\bibfnamefont{C.}~\bibnamefont{Giroud}},
  \bibinfo{author}{\bibfnamefont{M.~P.} \bibnamefont{Gryaznevich}},
  \bibinfo{author}{\bibfnamefont{D.}~\bibnamefont{Harting}},
  \bibnamefont{et~al.}, \bibinfo{journal}{Plasma Physics and Controlled Fusion}
  \textbf{\bibinfo{volume}{52}}, \bibinfo{pages}{105007}
  (\bibinfo{year}{2010}{\natexlab{a}}).

\bibitem[{\citenamefont{Sun et~al.}(2012)\citenamefont{Sun, Liang, Shaing, Liu,
  Koslowski, Jachmich, Alper, Alfier, Asunta, Buratti et~al.}}]{Sun_jet_textor}
\bibinfo{author}{\bibfnamefont{Y.}~\bibnamefont{Sun}},
  \bibinfo{author}{\bibfnamefont{Y.}~\bibnamefont{Liang}},
  \bibinfo{author}{\bibfnamefont{K.~C.} \bibnamefont{Shaing}},
  \bibinfo{author}{\bibfnamefont{Y.~Q.} \bibnamefont{Liu}},
  \bibinfo{author}{\bibfnamefont{H.~R.} \bibnamefont{Koslowski}},
  \bibinfo{author}{\bibfnamefont{S.}~\bibnamefont{Jachmich}},
  \bibinfo{author}{\bibfnamefont{B.}~\bibnamefont{Alper}},
  \bibinfo{author}{\bibfnamefont{A.}~\bibnamefont{Alfier}},
  \bibinfo{author}{\bibfnamefont{O.}~\bibnamefont{Asunta}},
  \bibinfo{author}{\bibfnamefont{P.}~\bibnamefont{Buratti}},
  \bibnamefont{et~al.}, \bibinfo{journal}{Nuclear Fusion}
  \textbf{\bibinfo{volume}{52}}, \bibinfo{pages}{083007}
  (\bibinfo{year}{2012}).

\bibitem[{\citenamefont{Rice et~al.}(2007)\citenamefont{Rice, Ince-Cushman,
  deGrassie, Eriksson, Sakamoto, Scarabosio, Bortolon, Burrell, Duval,
  Fenzi-Bonizec et~al.}}]{Rice}
\bibinfo{author}{\bibfnamefont{J.~E.} \bibnamefont{Rice}},
  \bibinfo{author}{\bibfnamefont{A.}~\bibnamefont{Ince-Cushman}},
  \bibinfo{author}{\bibfnamefont{J.~S.} \bibnamefont{deGrassie}},
  \bibinfo{author}{\bibfnamefont{L.-G.} \bibnamefont{Eriksson}},
  \bibinfo{author}{\bibfnamefont{Y.}~\bibnamefont{Sakamoto}},
  \bibinfo{author}{\bibfnamefont{A.}~\bibnamefont{Scarabosio}},
  \bibinfo{author}{\bibfnamefont{A.}~\bibnamefont{Bortolon}},
  \bibinfo{author}{\bibfnamefont{K.~H.} \bibnamefont{Burrell}},
  \bibinfo{author}{\bibfnamefont{B.~P.} \bibnamefont{Duval}},
  \bibinfo{author}{\bibfnamefont{C.}~\bibnamefont{Fenzi-Bonizec}},
  \bibnamefont{et~al.}, \bibinfo{journal}{Nuclear Fusion}
  \textbf{\bibinfo{volume}{47}}, \bibinfo{pages}{1618} (\bibinfo{year}{2007}).

\bibitem[{\citenamefont{Shaing et~al.}(2008)\citenamefont{Shaing, Cahyna,
  Becoulet, Park, Sabbagh, and Chu}}]{Shaing08}
\bibinfo{author}{\bibfnamefont{K.~C.} \bibnamefont{Shaing}},
  \bibinfo{author}{\bibfnamefont{P.}~\bibnamefont{Cahyna}},
  \bibinfo{author}{\bibfnamefont{M.}~\bibnamefont{Becoulet}},
  \bibinfo{author}{\bibfnamefont{J.-K.} \bibnamefont{Park}},
  \bibinfo{author}{\bibfnamefont{S.~A.} \bibnamefont{Sabbagh}},
  \bibnamefont{and} \bibinfo{author}{\bibfnamefont{M.~S.} \bibnamefont{Chu}},
  \bibinfo{journal}{Physics of Plasmas} \textbf{\bibinfo{volume}{15}},
  \bibinfo{eid}{082506} (\bibinfo{year}{2008}).

\bibitem[{\citenamefont{Shaing et~al.}(2009{\natexlab{a}})\citenamefont{Shaing,
  Sabbagh, and Chu}}]{Shaing09_1}
\bibinfo{author}{\bibfnamefont{K.~C.} \bibnamefont{Shaing}},
  \bibinfo{author}{\bibfnamefont{S.~A.} \bibnamefont{Sabbagh}},
  \bibnamefont{and} \bibinfo{author}{\bibfnamefont{M.~S.} \bibnamefont{Chu}},
  \bibinfo{journal}{Plasma Physics and Controlled Fusion}
  \textbf{\bibinfo{volume}{51}}, \bibinfo{pages}{035004}
  (\bibinfo{year}{2009}{\natexlab{a}}).

\bibitem[{\citenamefont{Shaing et~al.}(2009{\natexlab{b}})\citenamefont{Shaing,
  Sabbagh, and Chu}}]{Shaing09_2}
\bibinfo{author}{\bibfnamefont{K.~C.} \bibnamefont{Shaing}},
  \bibinfo{author}{\bibfnamefont{S.~A.} \bibnamefont{Sabbagh}},
  \bibnamefont{and} \bibinfo{author}{\bibfnamefont{M.~S.} \bibnamefont{Chu}},
  \bibinfo{journal}{Plasma Physics and Controlled Fusion}
  \textbf{\bibinfo{volume}{51}}, \bibinfo{pages}{035009}
  (\bibinfo{year}{2009}{\natexlab{b}}).

\bibitem[{\citenamefont{Shaing et~al.}(2009{\natexlab{c}})\citenamefont{Shaing,
  Sabbagh, and Chu}}]{Shaing09_3}
\bibinfo{author}{\bibfnamefont{K.~C.} \bibnamefont{Shaing}},
  \bibinfo{author}{\bibfnamefont{S.~A.} \bibnamefont{Sabbagh}},
  \bibnamefont{and} \bibinfo{author}{\bibfnamefont{M.~S.} \bibnamefont{Chu}},
  \bibinfo{journal}{Plasma Physics and Controlled Fusion}
  \textbf{\bibinfo{volume}{51}}, \bibinfo{pages}{055003}
  (\bibinfo{year}{2009}{\natexlab{c}}).

\bibitem[{\citenamefont{Shaing et~al.}(2009{\natexlab{d}})\citenamefont{Shaing,
  Chu, and Sabbagh}}]{Shaing09_4}
\bibinfo{author}{\bibfnamefont{K.~C.} \bibnamefont{Shaing}},
  \bibinfo{author}{\bibfnamefont{M.~S.} \bibnamefont{Chu}}, \bibnamefont{and}
  \bibinfo{author}{\bibfnamefont{S.~A.} \bibnamefont{Sabbagh}},
  \bibinfo{journal}{Plasma Physics and Controlled Fusion}
  \textbf{\bibinfo{volume}{51}}, \bibinfo{pages}{075015}
  (\bibinfo{year}{2009}{\natexlab{d}}).

\bibitem[{\citenamefont{Shaing et~al.}(2010)\citenamefont{Shaing, Sabbagh, and
  Chu}}]{Shaing10}
\bibinfo{author}{\bibfnamefont{K.~C.} \bibnamefont{Shaing}},
  \bibinfo{author}{\bibfnamefont{S.~A.} \bibnamefont{Sabbagh}},
  \bibnamefont{and} \bibinfo{author}{\bibfnamefont{M.~S.} \bibnamefont{Chu}},
  \bibinfo{journal}{Nuclear Fusion} \textbf{\bibinfo{volume}{50}},
  \bibinfo{pages}{025022} (\bibinfo{year}{2010}).

\bibitem[{\citenamefont{Kim et~al.}(2012)\citenamefont{Kim, Park, Kramer, and
  Boozer}}]{Kim_2012_POP}
\bibinfo{author}{\bibfnamefont{K.}~\bibnamefont{Kim}},
  \bibinfo{author}{\bibfnamefont{J.-K.} \bibnamefont{Park}},
  \bibinfo{author}{\bibfnamefont{G.~J.} \bibnamefont{Kramer}},
  \bibnamefont{and} \bibinfo{author}{\bibfnamefont{A.~H.}
  \bibnamefont{Boozer}}, \bibinfo{journal}{Physics of Plasmas}
  \textbf{\bibinfo{volume}{19}}, \bibinfo{pages}{082503}
  (\bibinfo{year}{2012}).

\bibitem[{\citenamefont{Logan et~al.}(2013)\citenamefont{Logan, Park, Kim,
  Wang, and Berkery}}]{Logan_2013_POP}
\bibinfo{author}{\bibfnamefont{N.~C.} \bibnamefont{Logan}},
  \bibinfo{author}{\bibfnamefont{J.-K.} \bibnamefont{Park}},
  \bibinfo{author}{\bibfnamefont{K.}~\bibnamefont{Kim}},
  \bibinfo{author}{\bibfnamefont{Z.}~\bibnamefont{Wang}}, \bibnamefont{and}
  \bibinfo{author}{\bibfnamefont{J.~W.} \bibnamefont{Berkery}},
  \bibinfo{journal}{Physics of Plasmas} \textbf{\bibinfo{volume}{20}},
  \bibinfo{pages}{122507} (\bibinfo{year}{2013}).

\bibitem[{\citenamefont{Liu et~al.}(2013)\citenamefont{Liu, Kirk, and
  Sun}}]{Liu_2013_POP}
\bibinfo{author}{\bibfnamefont{Y.~Q.} \bibnamefont{Liu}},
  \bibinfo{author}{\bibfnamefont{A.}~\bibnamefont{Kirk}}, \bibnamefont{and}
  \bibinfo{author}{\bibfnamefont{Y.}~\bibnamefont{Sun}},
  \bibinfo{journal}{Physics of Plasmas} \textbf{\bibinfo{volume}{20}},
  \bibinfo{pages}{042503} (\bibinfo{year}{2013}).

\bibitem[{\citenamefont{Wang et~al.}(2014)\citenamefont{Wang, Park, Liu, Logan,
  Kim, and Menard}}]{Wang_2014_POP}
\bibinfo{author}{\bibfnamefont{Z.}~\bibnamefont{Wang}},
  \bibinfo{author}{\bibfnamefont{J.-K.} \bibnamefont{Park}},
  \bibinfo{author}{\bibfnamefont{Y.~Q.} \bibnamefont{Liu}},
  \bibinfo{author}{\bibfnamefont{N.}~\bibnamefont{Logan}},
  \bibinfo{author}{\bibfnamefont{K.}~\bibnamefont{Kim}}, \bibnamefont{and}
  \bibinfo{author}{\bibfnamefont{J.~E.} \bibnamefont{Menard}},
  \bibinfo{journal}{Physics of Plasmas} \textbf{\bibinfo{volume}{21}},
  \bibinfo{pages}{042502} (\bibinfo{year}{2014}).

\bibitem[{\citenamefont{Satake et~al.}(2013)\citenamefont{Satake, Park, Sugama,
  and Kanno}}]{Satake_2013_NF}
\bibinfo{author}{\bibfnamefont{S.}~\bibnamefont{Satake}},
  \bibinfo{author}{\bibfnamefont{J.-K.} \bibnamefont{Park}},
  \bibinfo{author}{\bibfnamefont{H.}~\bibnamefont{Sugama}}, \bibnamefont{and}
  \bibinfo{author}{\bibfnamefont{R.}~\bibnamefont{Kanno}},
  \bibinfo{journal}{Nuclear Fusion} \textbf{\bibinfo{volume}{53}},
  \bibinfo{pages}{113033} (\bibinfo{year}{2013}).

\bibitem[{\citenamefont{Honda et~al.}(2014)\citenamefont{Honda, Satake, Suzuki,
  Matsunaga, Shinohara, Yoshida, Matsuyama, Ide, and Urano}}]{Honda_2014_NF}
\bibinfo{author}{\bibfnamefont{M.}~\bibnamefont{Honda}},
  \bibinfo{author}{\bibfnamefont{S.}~\bibnamefont{Satake}},
  \bibinfo{author}{\bibfnamefont{Y.}~\bibnamefont{Suzuki}},
  \bibinfo{author}{\bibfnamefont{G.}~\bibnamefont{Matsunaga}},
  \bibinfo{author}{\bibfnamefont{K.}~\bibnamefont{Shinohara}},
  \bibinfo{author}{\bibfnamefont{M.}~\bibnamefont{Yoshida}},
  \bibinfo{author}{\bibfnamefont{A.}~\bibnamefont{Matsuyama}},
  \bibinfo{author}{\bibfnamefont{S.}~\bibnamefont{Ide}}, \bibnamefont{and}
  \bibinfo{author}{\bibfnamefont{H.}~\bibnamefont{Urano}},
  \bibinfo{journal}{Nuclear Fusion} \textbf{\bibinfo{volume}{54}},
  \bibinfo{pages}{114005} (\bibinfo{year}{2014}).

\bibitem[{\citenamefont{Sovinec et~al.}(2004)\citenamefont{Sovinec, Glasser,
  Gianakon, Barnes, Nebel, Kruger, Schnack, Plimpton, Tarditi, and
  Chu}}]{Sovinec}
\bibinfo{author}{\bibfnamefont{C.~R.} \bibnamefont{Sovinec}},
  \bibinfo{author}{\bibfnamefont{A.~H.} \bibnamefont{Glasser}},
  \bibinfo{author}{\bibfnamefont{T.~A.} \bibnamefont{Gianakon}},
  \bibinfo{author}{\bibfnamefont{D.~C.} \bibnamefont{Barnes}},
  \bibinfo{author}{\bibfnamefont{R.~A.} \bibnamefont{Nebel}},
  \bibinfo{author}{\bibfnamefont{S.~E.} \bibnamefont{Kruger}},
  \bibinfo{author}{\bibfnamefont{D.~D.} \bibnamefont{Schnack}},
  \bibinfo{author}{\bibfnamefont{S.~J.} \bibnamefont{Plimpton}},
  \bibinfo{author}{\bibfnamefont{A.}~\bibnamefont{Tarditi}}, \bibnamefont{and}
  \bibinfo{author}{\bibfnamefont{M.~S.} \bibnamefont{Chu}},
  \bibinfo{journal}{Journal of Computational Physics}
  \textbf{\bibinfo{volume}{195}}, \bibinfo{pages}{355 } (\bibinfo{year}{2004}).

\bibitem[{\citenamefont{Sun et~al.}(2010{\natexlab{b}})\citenamefont{Sun,
  Liang, Shaing, Koslowski, Wiegmann, and Zhang}}]{Sun10_prl}
\bibinfo{author}{\bibfnamefont{Y.}~\bibnamefont{Sun}},
  \bibinfo{author}{\bibfnamefont{Y.}~\bibnamefont{Liang}},
  \bibinfo{author}{\bibfnamefont{K.~C.} \bibnamefont{Shaing}},
  \bibinfo{author}{\bibfnamefont{H.~R.} \bibnamefont{Koslowski}},
  \bibinfo{author}{\bibfnamefont{C.}~\bibnamefont{Wiegmann}}, \bibnamefont{and}
  \bibinfo{author}{\bibfnamefont{T.}~\bibnamefont{Zhang}},
  \bibinfo{journal}{Physical Review Letters} \textbf{\bibinfo{volume}{105}},
  \bibinfo{pages}{145002} (\bibinfo{year}{2010}{\natexlab{b}}).

\bibitem[{\citenamefont{Izzo and Joseph}(2008)}]{Izzo_rmp}
\bibinfo{author}{\bibfnamefont{V.~A.} \bibnamefont{Izzo}} \bibnamefont{and}
  \bibinfo{author}{\bibfnamefont{I.}~\bibnamefont{Joseph}},
  \bibinfo{journal}{Nuclear Fusion} \textbf{\bibinfo{volume}{48}},
  \bibinfo{pages}{115004} (\bibinfo{year}{2008}).

\bibitem[{\citenamefont{Zhu}(2012)}]{pzhu_aps}
\bibinfo{author}{\bibfnamefont{P.}~\bibnamefont{Zhu}},
  \bibinfo{journal}{Proceedings of 54th Annual Meeting of the APS Division of
  Plasma Physics} \textbf{\bibinfo{volume}{57}} (\bibinfo{year}{2012}).

\bibitem[{\citenamefont{Sun et~al.}(2011)\citenamefont{Sun, Liang, Shaing,
  Koslowski, Wiegmann, and Zhang}}]{Sun11_nf}
\bibinfo{author}{\bibfnamefont{Y.}~\bibnamefont{Sun}},
  \bibinfo{author}{\bibfnamefont{Y.}~\bibnamefont{Liang}},
  \bibinfo{author}{\bibfnamefont{K.~C.} \bibnamefont{Shaing}},
  \bibinfo{author}{\bibfnamefont{H.~R.} \bibnamefont{Koslowski}},
  \bibinfo{author}{\bibfnamefont{C.}~\bibnamefont{Wiegmann}}, \bibnamefont{and}
  \bibinfo{author}{\bibfnamefont{T.}~\bibnamefont{Zhang}},
  \bibinfo{journal}{Nuclear Fusion} \textbf{\bibinfo{volume}{51}},
  \bibinfo{pages}{053015} (\bibinfo{year}{2011}).

\bibitem[{\citenamefont{Sun et~al.}(2013)\citenamefont{Sun, Shaing, Liang,
  Shen, and Wan}}]{Sun_2013_NF}
\bibinfo{author}{\bibfnamefont{Y.}~\bibnamefont{Sun}},
  \bibinfo{author}{\bibfnamefont{K.~C.} \bibnamefont{Shaing}},
  \bibinfo{author}{\bibfnamefont{Y.}~\bibnamefont{Liang}},
  \bibinfo{author}{\bibfnamefont{B.}~\bibnamefont{Shen}}, \bibnamefont{and}
  \bibinfo{author}{\bibfnamefont{B.}~\bibnamefont{Wan}},
  \bibinfo{journal}{Nuclear Fusion} \textbf{\bibinfo{volume}{53}},
  \bibinfo{pages}{073026} (\bibinfo{year}{2013}).

\bibitem[{\citenamefont{Boozer}(2006)}]{Boozer_2006_POP}
\bibinfo{author}{\bibfnamefont{A.~H.} \bibnamefont{Boozer}},
  \bibinfo{journal}{Physics of Plasmas} \textbf{\bibinfo{volume}{13}},
  \bibinfo{pages}{044501} (\bibinfo{year}{2006}).

\bibitem[{\citenamefont{Park et~al.}(2009{\natexlab{b}})\citenamefont{Park,
  Boozer, Menard, Garofalo, Schaffer, Hawryluk, Kaye, Gerhardt, and
  Sabbagh}}]{Park_2009_POP}
\bibinfo{author}{\bibfnamefont{J.-K.} \bibnamefont{Park}},
  \bibinfo{author}{\bibfnamefont{A.~H.} \bibnamefont{Boozer}},
  \bibinfo{author}{\bibfnamefont{J.~E.} \bibnamefont{Menard}},
  \bibinfo{author}{\bibfnamefont{A.~M.} \bibnamefont{Garofalo}},
  \bibinfo{author}{\bibfnamefont{M.~J.} \bibnamefont{Schaffer}},
  \bibinfo{author}{\bibfnamefont{R.~J.} \bibnamefont{Hawryluk}},
  \bibinfo{author}{\bibfnamefont{S.~M.} \bibnamefont{Kaye}},
  \bibinfo{author}{\bibfnamefont{S.~P.} \bibnamefont{Gerhardt}},
  \bibnamefont{and} \bibinfo{author}{\bibfnamefont{S.~A.}
  \bibnamefont{Sabbagh}}, \bibinfo{journal}{Physics of Plasmas}
  \textbf{\bibinfo{volume}{16}}, \bibinfo{pages}{056115}
  (\bibinfo{year}{2009}{\natexlab{b}}).

\bibitem[{\citenamefont{Sun et~al.}(2015)\citenamefont{Sun, Liang, Qian, Shen,
  and Wan}}]{Sun_2015_PPCF}
\bibinfo{author}{\bibfnamefont{Y.}~\bibnamefont{Sun}},
  \bibinfo{author}{\bibfnamefont{Y.}~\bibnamefont{Liang}},
  \bibinfo{author}{\bibfnamefont{J.}~\bibnamefont{Qian}},
  \bibinfo{author}{\bibfnamefont{B.}~\bibnamefont{Shen}}, \bibnamefont{and}
  \bibinfo{author}{\bibfnamefont{B.}~\bibnamefont{Wan}},
  \bibinfo{journal}{Plasma Physics and Controlled Fusion}
  \textbf{\bibinfo{volume}{57}}, \bibinfo{pages}{045003}
  (\bibinfo{year}{2015}).

\bibitem[{\citenamefont{Politzer et~al.}(2008)\citenamefont{Politzer, Petty,
  Jayakumar, Luce, Wade, DeBoo, Ferron, Gohil, Holcomb, Hyatt
  et~al.}}]{DIII-D_NBI}
\bibinfo{author}{\bibfnamefont{P.~A.} \bibnamefont{Politzer}},
  \bibinfo{author}{\bibfnamefont{C.~C.} \bibnamefont{Petty}},
  \bibinfo{author}{\bibfnamefont{R.~J.} \bibnamefont{Jayakumar}},
  \bibinfo{author}{\bibfnamefont{T.~C.} \bibnamefont{Luce}},
  \bibinfo{author}{\bibfnamefont{M.~R.} \bibnamefont{Wade}},
  \bibinfo{author}{\bibfnamefont{J.~C.} \bibnamefont{DeBoo}},
  \bibinfo{author}{\bibfnamefont{J.~R.} \bibnamefont{Ferron}},
  \bibinfo{author}{\bibfnamefont{P.}~\bibnamefont{Gohil}},
  \bibinfo{author}{\bibfnamefont{C.~T.} \bibnamefont{Holcomb}},
  \bibinfo{author}{\bibfnamefont{A.~W.} \bibnamefont{Hyatt}},
  \bibnamefont{et~al.}, \bibinfo{journal}{Nuclear Fusion}
  \textbf{\bibinfo{volume}{48}}, \bibinfo{pages}{075001}
  (\bibinfo{year}{2008}).

\end{thebibliography}

\newpage
\clearpage

\newpage
\clearpage

\newpage
\clearpage

\newpage
\clearpage

\newpage
\clearpage

\newpage
\clearpage

\newpage
\clearpage

\newpage
\clearpage

\newpage
\clearpage

\newpage
\clearpage

\end{document}